\documentclass[aps,prb,reprint,superscriptaddress]{revtex4-1}

\usepackage{color}
\usepackage{graphicx}  
\usepackage{dcolumn}   
\usepackage{bm}        
\usepackage{amssymb}   
\newcommand{\Co}{Sc$_3$CoC$_4$}

\newcommand{\issrT}{$I_{XRD}(T)$\ }
\newcommand{\rhoT}{$\rho(T)$}

\begin{document}


\title{Evidence for a soft-phonon-mode-driven
  Peierls-type distortion in \Co}


\author{Jan Langmann}
\author{Christof Haas}
\affiliation{University of Augsburg, Institut f\"ur Physik,
  Universit\"at Augsburg,
  Universit\"atsstra\ss e 1, D-86159 Augsburg, Germany}
\author{Emmanuel  Wenger}
\author{Dominik Schaniel}
\affiliation{Université de Lorraine, CNRS, CRM2, F-54000 Nancy, France}
\author{Wolfgang Scherer}
\author{Georg Eickerling}
\email{georg.eickerling@physik.uni-augsburg.de}
\affiliation{University of Augsburg, Institut f\"ur Physik,
  Universit\"at Augsburg,
  Universit\"atsstra\ss e 1, D-86159 Augsburg, Germany}


\date{\today}

\begin{abstract}
  We provide experimental and theoretical evidence for the realization
  of the Peierls-type structurally distorted state in the quasi-one-dimensional
  superconductor \Co\ by a phonon-softening mechanism. The transition
  from the high- to the final low-temperature phase below 80~K
  proceeds \textit{via} an extended intermediate temperature regime
  between 80~K and 150~K characterized by phonon--driven atom displacements.
  In support of the low-dimensional character of the title compound we
  find a highly anisotropic correlation-length of these dynamic distortions.
\end{abstract}


\maketitle


\section{Introduction}\label{sec:intro}

Due to recent advances in nanofabrication and nanopatterning
dimensionality effects in physical phenomena have become a very active
field of research.  Even nanowires or two-dimensional films of
decidedly three-dimensional materials such as perovskites have become
available for further studies.\cite{Schlaus19, Ji19}  Still, the
majority of publications in this field is dedicated to a narrow range
of materials with structurally inherent low-dimensional
features. These are for example NbSe$_3$ \cite{Meerschaut75, Hodeau78,
  Regueiro92, Yang19} and K$_{0.3}$MoO$_3$ \cite{Graham66,
  Travaglini81, Pouget83a, Inagaki18} as representatives of
quasi-one-dimensional (quasi-1D) compounds and graphite/graphene
\cite{Novoselov04, Cao18} and transition-metal dichalcogenides
\cite{Williams74, Lee19, Hughes77, Shu19, Pyon12, Oike18} as
representatives of quasi-two-dimensional (quasi-2D) compounds. On the
one hand, this focus comes from the weak bonding between their
low-dimensional building units easing the fabrication of
nano-devices. On the other hand, but even more importantly,
intriguing effects of the strongly anisotropic atomic interactions can
already be observed in the bulk material.

A characteristic physical phenomenon in many structurally
low-dimensional materials is the existence of a subtle competition
between a structurally distorted state (\textit{e.g.} due to the formation
of a charge-density wave) and a superconducting state. The balance
between these usually conflicting states may be influenced by external
factors such as hydrostatic pressure,\cite{Ido90, Regueiro92,
  Yasuzuka05, Monceau12, Kiswandhi13} rapid quenching \cite{Oike18}
and chemical pressure \textit{via} the intercalation or substitution
of additional elements.\cite{Yang12, Pyon12}

The complex carbide \Co\ represents a promising new member in this
family of materials: Its structure is coined by quasi-1D infinite
$\left[\right.$Co(C$_2$)$_2\left.\right]$ ribbons orientated along the
crystallographic $b$-axis
\cite{jeitschko_carbon_1989,Tsokol86,Rohrmoser07, Scherer10,
  Scheidt11, Scherer12, Eickerling13, He15} and it shows a phase
transition into a superconducting state below
$T_{\mathtt{c}} =$~4.5~K.\cite{Scheidt11,Scherer10,Eickerling13} As was
recently demonstrated by Wang \textit{et al.}, the superconducting
volume fraction of polycrystalline \Co\ samples significantly
increases with pressure.\cite{Wang16} At the same time, we have shown
in a previous combined X-ray and neutron diffraction study that below
approx. 72~K \Co\ undergoes a Peierls-type transition to a
low-temperature phase with a doubled translational period along the
$\left[\right.$Co(C$_2$)$_2\left.\right]$ ribbons.  \cite{Scherer10,
  Scherer12, Eickerling13} Crystallographically, the transition from
the orthorhombic high-temperature (HT) phase (space group $Immm$) to
the monoclinic low-temperature (LT) phase (space group $C2/m$) proceeds
\textit{via} a $t2$- followed by an $i2$-transition, leading to a
systematic twinning of single crystalline samples in the LT
phase.\cite{Vogt09} Yet, the driving forces and the exact path to this
structurally distorted LT phase remain controversial.  In earlier
works, we interpreted anomalies in the electrical resistivity and the
magnetic susceptibility of polycrystalline \Co\ samples as hints to
the emergence of a charge-density wave (CDW) below
$\approx$~140~K. \cite{Eickerling13, Scherer10, Scherer12} This
interpretation has been challenged by Zhang \textit{et al.}
\cite{Zhang12} Their theoretical study provided no evidence of a Fermi
surface instability with respect to a CDW of \Co\ in its HT phase. In
order to correlate the anomalies in the electrical transport
properties with potential structural changes we performed
temperature-dependent X-ray diffraction and resistivity measurements
on single-crystalline \Co\ samples.

\section{Methods}\label{sec:experimental}

Single crystals of \Co\ were grown according to methods described in
the literature\cite{Vogt09, Rohrmoser07, He15} and in addition from a
lithium metal flux.\cite{Haas19} Needle-like samples with a thickness
of $\approx 20$~$\mu$m and a length of $\approx 200$~$\mu$m were
obtained from the first method and platelet-like samples with a
thickness of $\approx 150$~$\mu$m and a lateral size of
$\approx 300$~$\mu$m from the latter.  X-ray diffraction
measurements\footnote{See Supplemental Material at [URL will be
  inserted by publisher] for full experimental details and all data
  recorded between room temperature and 12~K.} of the temperature
dependent superstructure reflection intensities (10~K~$< T <$~160~K)
were performed on a HUBER eulerian cradle equipped with a MAR345
image-plate detector and operated at the window of a BRUKER FR591
rotating anode (Mo K$_{\alpha}$). Additional temperature-dependent
single crystal data ($T > 100$~K) was collected on a BRUKER-NONIUS
$\kappa$-goniometer operated at the window of an INCOATEC MicroFocus
Tube (Mo K$_{\alpha}$). Mapping of diffuse scattering intensities was
done employing a HUBER eulerian cradle goniometer equipped with an
INCOATEC MicroFocus Tube (Ag K$_{\alpha}$) and a PILATUS 300K CdTe
detector. Cryogenic temperatures $T >$~80~K were generated by a
standard OXFORD open-flow N$_2$ cooler,\cite{Cosier86} measurements at
$T <$~80~K were performed employing an ARS closed-cycle
He-cryostat. The handling of parasitic scattering from the vacuum and
radiation shields in the closed-cycle cryostat (beryllium domes) was
described elsewhere.\cite{Reisinger07} Numerical values for the
intensities of representative superstructure reflections and diffuse
features were extracted by image analysis of the collected X-ray
diffraction data.

Resistivity measurements of single crystals of \Co\ contacted
in a four-point geometry were carried out
using a Physical Property Measurement System (PPMS, QUANTUM DESIGN).

Geometry relaxations of the HT and LT phase of \Co\ starting from the
structural parameters from Eickerling \textit{et
  al.}\cite{Eickerling13} were performed employing the VASP
code.\cite{kresse_efficiency_1996,kresse_efficient_1996,kresse_ab_1994,kresse_ab_1993}
The PBE density functional was used throughout,\cite{perdew96,
  perdew97} the energy cutoff for the plane wave basis set was set to
550~eV and a Brillouin grid sampling of $4~\times 4~\times 2$ and
$2~\times 2~\times 2$ was used for the HT- and LT-phase,
respectively. Optimizations were stopped when forces were smaller than
0.001~eV/\AA.

The PHONOPY code\cite{Togo15} was used for phonon dispersion
calculations on \Co\ with a $2~\times 2~\times 2$ supercell. Forces
were calculated with the VASP program employing the PBE density
functional\cite{perdew96, perdew97}, a $4~\times 4~\times 2$ $k$-point
mesh and an energy cutoff for the plane wave basis set of
500~eV.\cite{kresse_efficiency_1996,kresse_efficient_1996,kresse_ab_1994,kresse_ab_1993}
Temperature-dependent thermal diffuse scattering (TDS) simulations
were obtained using the \textit{ab2tds} code.\cite{Wehinger13} The
phonon eigenvectors underlying the simulations were generated with
PHONOPY on a $24~\times 24~\times 22$ $q-$mesh (containing
$\Gamma$). In \textit{ab2tds}, the Fourier transform of the dynamical
matrix was calculated on a $9~\times 9~\times 9$ mesh of
points. Debye-Waller factors for each temperature were computed and
reciprocal space planes of TDS intensity were sampled on 100~$\times$
100 $q$-points for a wavelength of 0.56087~\AA\ and a lower
eigenvalue-cutoff of 0.001~meV.

\section{Electrical Resistivity}\label{sec:res}
\begin{figure}
   \includegraphics[height=0.58\textwidth]{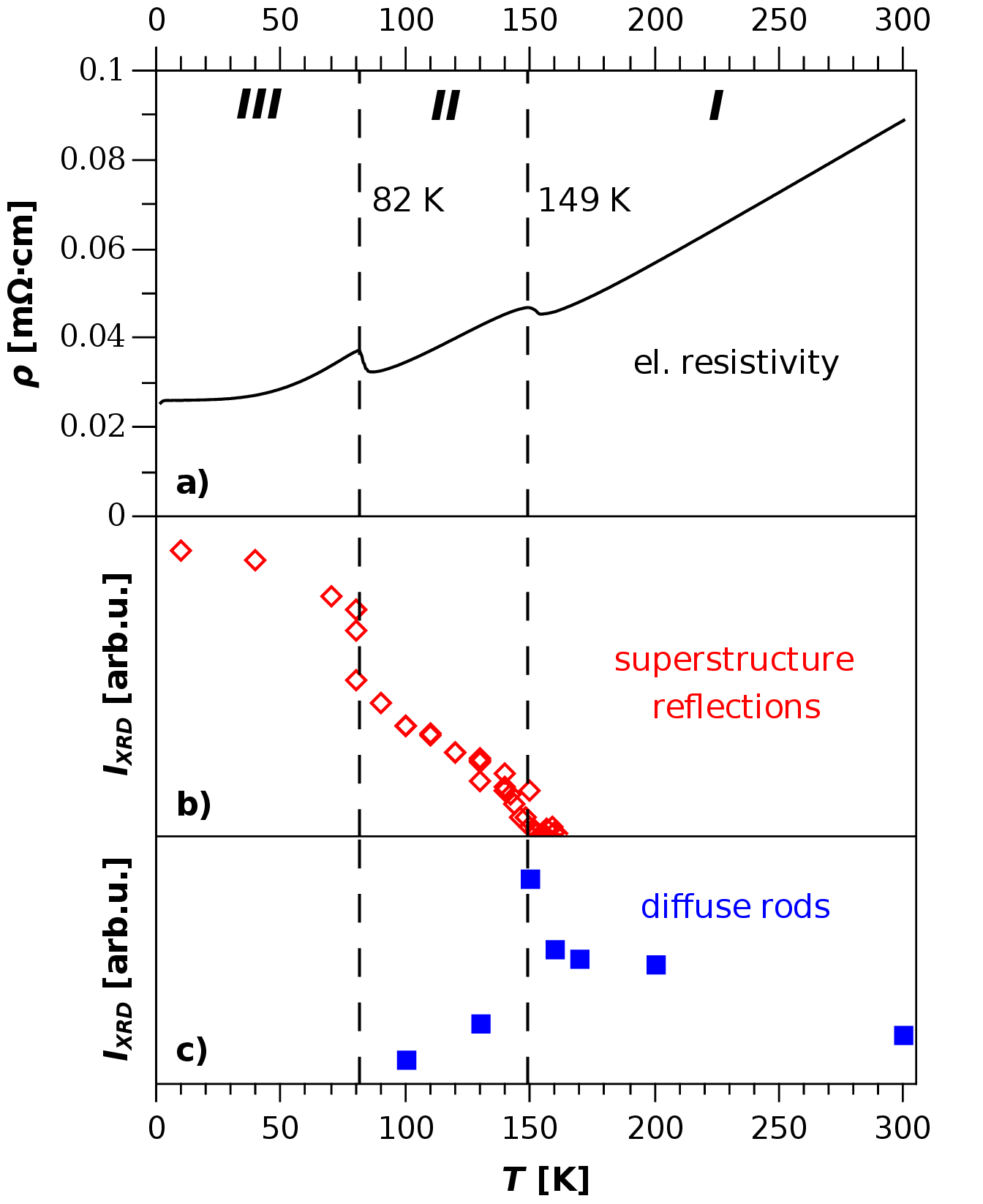}
   \caption{\label{fig:res-xrd} Temperature-dependence of (a) the
     electrical resistivity \rhoT, (b) the X-ray scattering intensity
     of superstructure reflections and (c) the X-ray scattering
     intensity of diffuse rods \issrT\ connecting the superstructure
     reflection positions along $c^\ast$.}
  \end{figure}

We focus on the results of the electrical resistivity measurements on
single crystalline samples of \Co\ first (see
Fig.~\ref{fig:res-xrd}a). In accordance with earlier resistivity data
from polycrystalline samples,\cite{Eickerling13, Scherer10} two
anomalies are observed at $T \approx$~82~K and 149~K. At both
temperatures, the overall metallic decrease of \rhoT\ is interrupted
by a local increase of the resistivity. The anomaly at 82~K is
discerned by a sharp jump in \rhoT\ and corresponds to an
irreversibility in \rhoT\ for polycrystalline samples that permanently
displaces the heating against the cooling curve.\cite{Eickerling13,
  Scherer10} This contrasts with the broader and more gradual
character of the anomaly at 149~K.

Details about structural changes connected to the two anomalies in the
electrical resistivity are provided by the results of detailed
temperature-dependent single-crystal X-ray diffraction measurements on
\Co\ outlined in the following.

\section{X-ray diffraction}\label{sec:recspace-map}

A concise analysis of temperature-dependent changes in diffraction
space allows insight into the evolution of the LT structure of \Co\
from the HT structure.  The intensity of the superstructure
reflections \issrT with
$k = \left(\pm \frac{1}{2}, \pm \frac{1}{2}, 0\right)$ that result
from a fourfold enlargement of the orthorhombic HT unit cell in its
$ab$-plane \cite{Vogt09} represents an appropriate order parameter
for the transition from the HT to the LT structure. The simultaneous
existence of additional reflections at
$k = \left(+\frac{1}{2}, +\frac{1}{2}, 0\right)$ and
$k = \left(+\frac{1}{2}, -\frac{1}{2}, 0\right)$ is due to the
systematic twinning caused by the $t2$-transition. Note, that all real
space and reciprocal space coordinates given hereafter are referred to
the orthorhombic HT-phase unit cell.

\begin{figure}[h]
  \centering
   \includegraphics[height=0.56\textwidth]{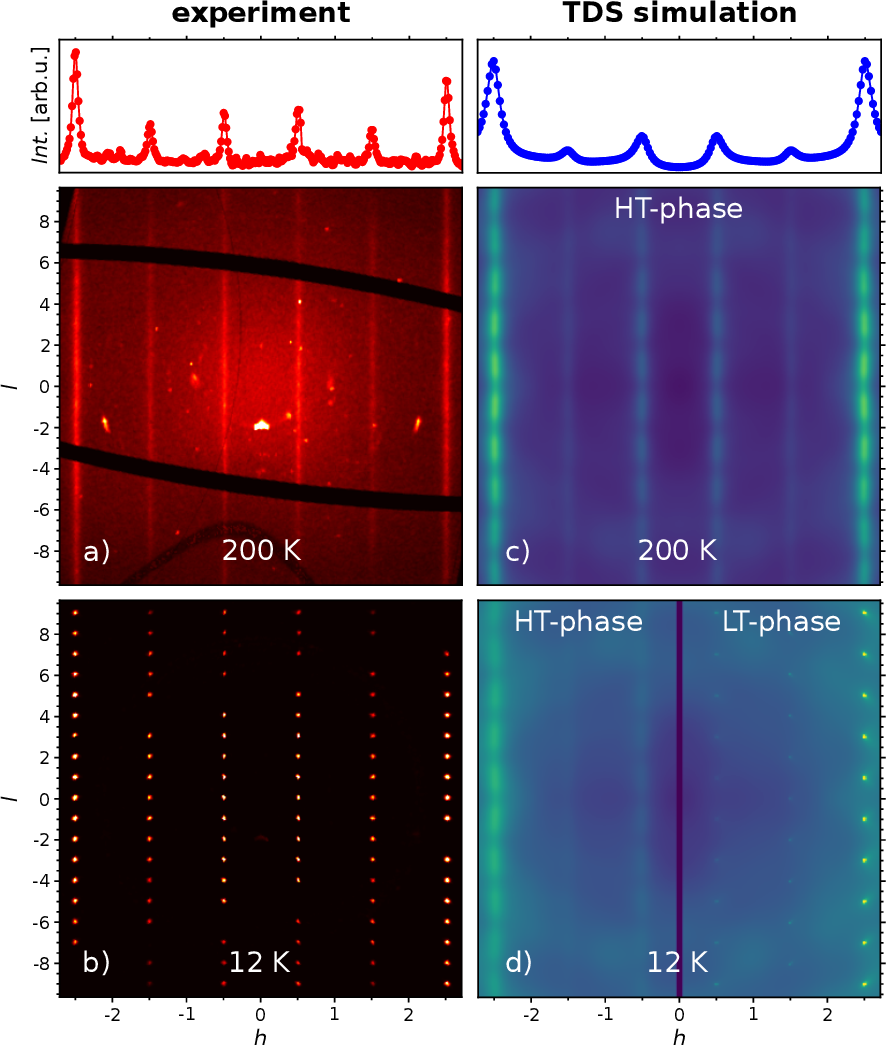}
   \caption{Comparison of the X-ray scattering features in the
     $(h, 1.5, l)$-plane of \Co\ as obtained from experiments at
     different temperatures (a and b) and thermal diffuse scattering
     (TDS) simulations based on \textit{ab-initio} calculated phonon
     dispersion relations for the HT-phase structure (c and left part
     of d) and the LT-phase structure of \Co (right part of d). Note
     that the Miller indices refer to the orthorhombic HT-phase and
     that twinning was not considered in the calculations. 1D-profiles
     of (a) and (c) at $l \approx 2$ are given in the top panel. For
     details on the simulations, see main text.}
  \label{fig:tds-sim}
\end{figure}

Our measurements reveal an increase of \issrT (see
Fig.~\ref{fig:res-xrd}b) in two phenomenologically distinct
steps at temperatures between 150~K and 80~K.\footnote{An account of
  the employed image analysis techniques for the extraction of \issrT
  from experimental X-ray diffraction data is given in the
  Supplemental Material at [URL will be inserted by
  publisher]} Thereby, $T\approx 150~K$ marks the onset of \issrT
followed by a steady increase down to 80~K. At about 80~K, a sharp
jump of \issrT is observed. Further cooling towards 10~K entails a
saturation of \issrT already below $\approx$~70~K. We note the close
resemblance of this temperature-dependence of the superstructure
reflection intensity to the observed behavior of \issrT in the
charge-density wave material $2H$-TaSe$_2$.\cite{Moncton75, Moncton77,
  Williams76} In this compound, a sharp step in the superstructure
reflection intensities marks a lock-in transition from an
incommensurate modulation of the atomic positions at higher
temperatures to a commensurate modulation at lower temperatures
\cite{Moncton75, Moncton77, Williams76}.  However, within the
available experimental accuracy we could not find hints to the
existence of an incommensurate phase in \Co, \textit{i.e.} significant
temperature-dependent changes in the superstructure reflection
positions or the appearance of higher-order satellite
reflections. This puts \Co\ in a row with the extensively studied
transition-metal dichalcogenide $1T$-TiSe$_2$ that shows a
Peierls-type structural distortion with a twofold commensurable
modulation wave vector $k$ down to a temperature of
8.3~K.\cite{Snow03,Kusmartseva09,Sugai80,
  Wakabayashi78,Hughes77,Holt01,DiSalvo76,Brown80,Kidd02,Joe14}

Adding to the pinpoint superstructure reflections strongly
temperature-dependent diffuse rods connecting the superstructure
reflection positions along $c^\ast$ can be observed for
\Co. Representative $(h, 1.5, l)$ reciprocal space planes
reconstructed from measuring data at 200~K and 12~K and showing
exclusively superstructure reflections and diffuse rods are shown in
Fig.~\ref{fig:tds-sim}a-b.\cite{Note1} Above 200~K, diffuse rod-shaped
features without significant intensity modulation along $c^\ast$ are
observed. Upon cooling towards 80~K, a monotonous increase of the
intensity at the superstructure reflection positions is paralleled by
a lambda-shaped peaking of the diffuse intensity between the
superstructure reflection positions at 150~K and its subsequent decay
to zero (see Fig.~\ref{fig:res-xrd}c).\cite{Note2}
Below 80~K, only intensity at the superstructure reflection positions
remains. This marked temperature-dependence along with an anomalous
modulation of the diffuse intensity with varying $h$-index (indicated
by the profile in the top-panel of Fig.~\ref{fig:tds-sim}a) rules out
crystal defects (\textit{e.g.} stacking disorder) as the predominant
origin of these diffuse features. A detailed discussion of the
characteristic variation of the diffuse intensity in reciprocal space
can be found in Appendix~B.

Moreover, the coinciding positions of both,
diffuse rods and pinpoint superstructure reflections, can be taken as
a hint to their common origin. Similar transitions from precursor
diffuse features in reciprocal space to pinpoint superstructure
reflections are known for other structurally low-dimensional materials
featuring low-temperature periodic distortions, \textit{e.g.} $1T$-TaS$_2$,
\cite{Williams74, Scruby75} K$_{0.3}$MoO$_3$\cite{Pouget83a,
  Pouget16} or NbSe$_3$.\cite{Pouget83, Pouget16}

Based on the temperature-dependent changes in diffraction space, three
different temperature regimes for the structural properties of \Co\
may be assigned: (\textit{\textbf{I}})~a HT-regime above
$\approx$~150~K characterized by unmodulated (or only weakly)
modulated diffuse rods along $c^\ast$ in reciprocal space,
(\textit{\textbf{II}})~a pre-LT-regime between $\approx$~150~K and
$\approx$~80~K with coexistent diffuse rods and weak superstructure
reflections, and (\textit{\textbf{III}})~a LT-regime below
$\approx$~80~K marked by the exclusive presence of strong and pinpoint
superstructure reflections. This partitioning into temperature regimes
fits equally well with the anomalies in the electrical resistivity
\rhoT\ (see Fig.~\ref{fig:res-xrd}a).  More specifically, the steady
transition between (\textit{\textbf{I}}) and (\textit{\textbf{II}}) in
\issrT is reflected by a broad uprise in \rhoT\ at 149~K. At the same
time, the step-like increase of \issrT between (\textit{\textbf{II}})
and (\textit{\textbf{III}}) relates to the sharp increase in \rhoT\ at
82~K. The differing nature of the transitions at around ~80~K and
150~K is further emphasized by powder neutron diffraction studies on
\Co\ performed earlier.\cite{Eickerling13} Therein, step-like lattice
parameter changes with sudden increases in $b$ and $c$ and a decrease
in $a$ were observed upon cooling of the samples to below 80~K. There
was no evidence for a comparable anomaly in $a$, $b$ and $c$ in the
temperature region around 150~K.

\section{Discussion}\label{sec:discussion}

\begin{figure*}
  \includegraphics[height=0.38\textwidth]{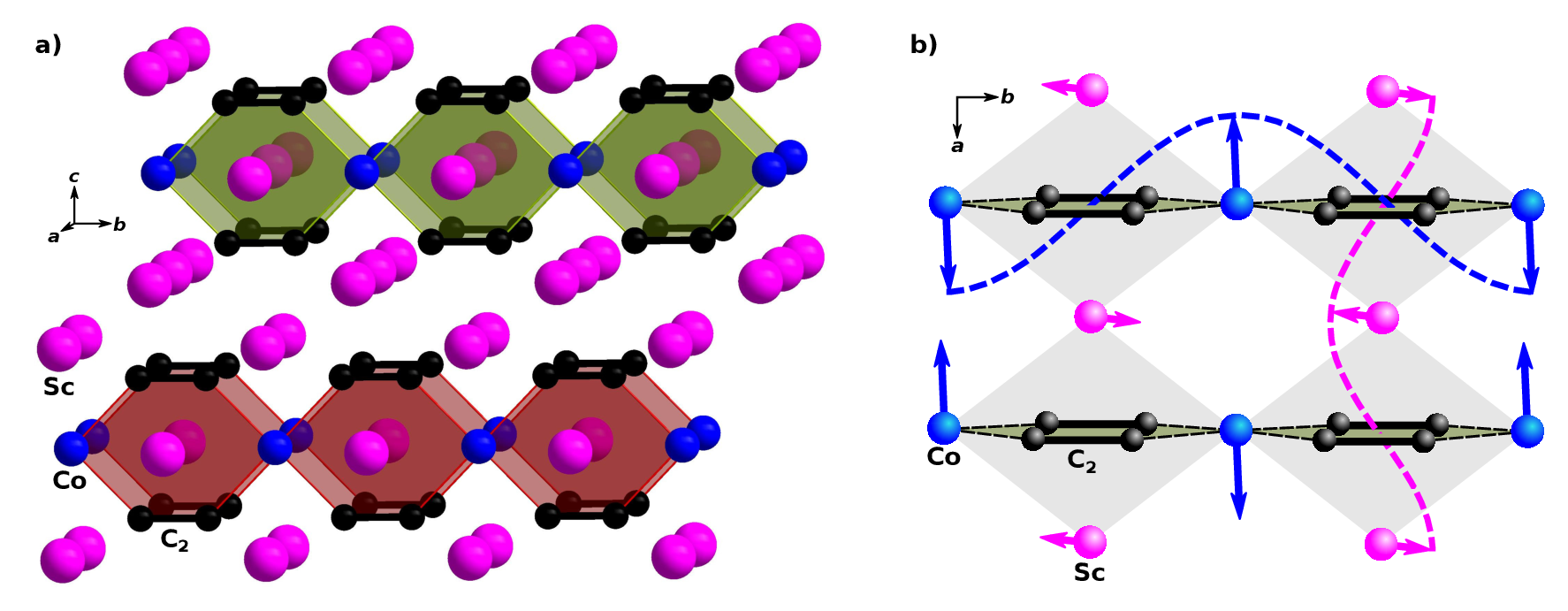}
  \caption{Ball-and-Stick representation of the layered building units
    of HT \Co\ in (a) the crystallographic $bc$-plane and (b) the
    crystallographic $ab$-plane of the orthorhombic unit cell. In (b) the
    sinusoidal displacive modulation of the cobalt and scandium atom positions
    as observed for the low-frequency phonon modes between W and T
    (see text) and the monoclinic LT-phase\cite{Eickerling13} is
    indicated by arrows.}
  \label{fig:struct-layers}
\end{figure*}

Starting point for the interpretation of the above results is the
structural model of the \Co\ HT-phase. The existence of rod-shaped
features in diffraction space may be related to layered structural moieties
in real space. Taking into account the orientation of the rods along
$c^\ast$, these layers must extend parallel to the crystallographic
$ab$-plane of the orthorhombic HT cell and can be associated
with stacked ribbons of interconnected
$\left[\right.$Co(C$_2$)$_2$Co$\left.\right]$ hexagons (shaded in
red and green in Fig.~\ref{fig:struct-layers}a).

In a simplistic picture, the diffuse rods may be attributed to
disorder between the layered building units of the HT-structure along
$c$. However, the characteristic intensity modulation of the rod
intensity perpendicular to $c^\ast$ (see profile above
Fig.~\ref{fig:tds-sim}a) precludes an explanation in terms of a static
stacking disorder involving the slippage of complete layers (see
Appendix ~\ref{appendix2}). An alternative explanation for the
occurrence of diffuse intensity above 80~K might be provided by
precursor dynamic fluctuations along the displacement coordinates of
the static Peierls-type distortion evolving below 80~K. The
temperature-dependent contraction of the diffuse intensity into
superstructure reflections between 150~K and 80~K may in turn be
linked to the softening of a phonon mode at
$k = \left(\frac{1}{2}, \frac{1}{2}, 0\right)$.\cite{Eickerling13}

\begin{figure}[h]
  \centering
  \includegraphics[width=0.48\textwidth]{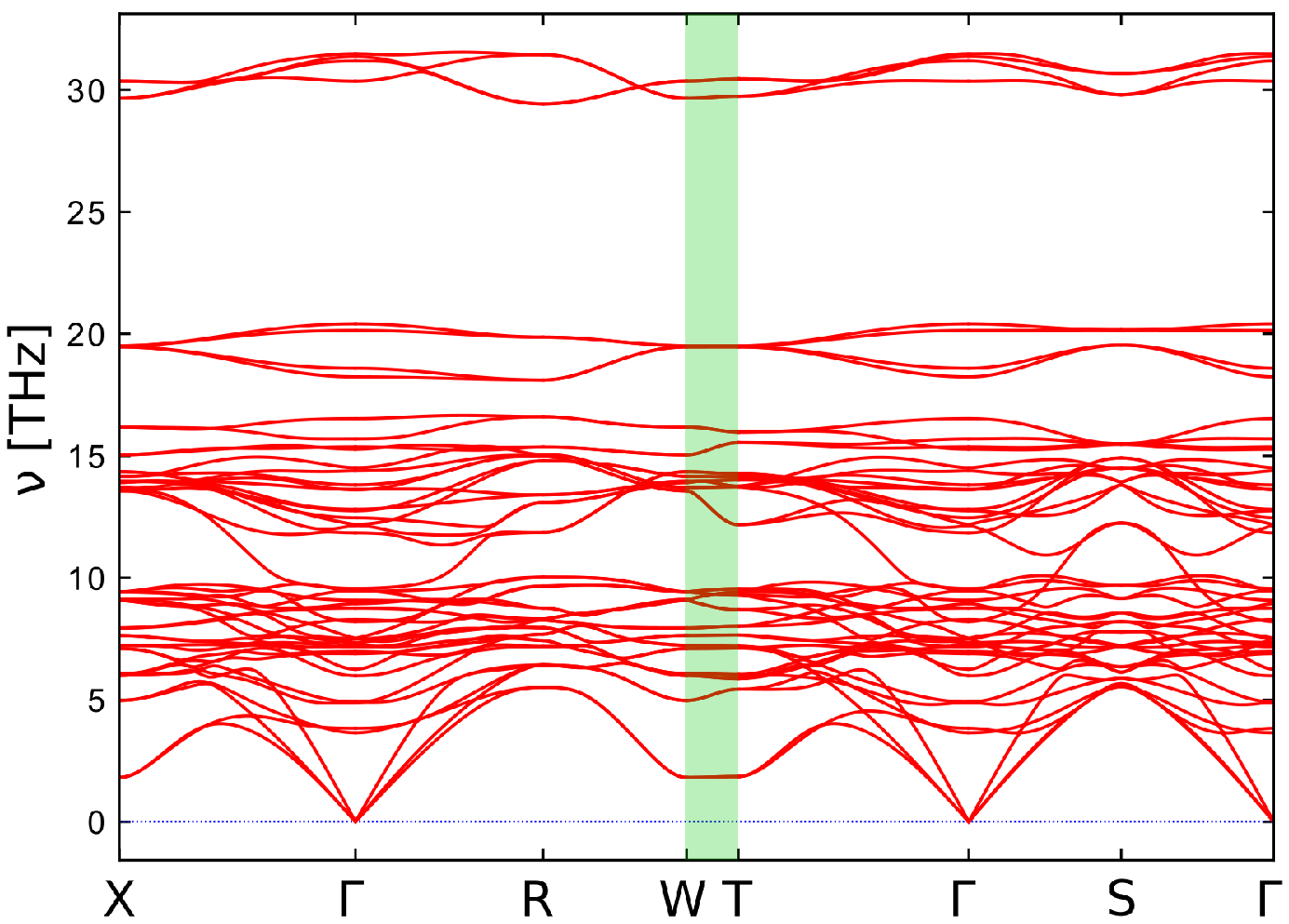}
  \caption{Calculated phonon dispersion along selected high-symmetry paths
    in the Brillouin zone of HT
    \Co. The path between W ($\frac{1}{2}, \frac{1}{2}, 0$) and
    T ($\frac{1}{2}, \frac{1}{2}, \frac{1}{2}$) is highlighted.}
  \label{fig:ht-phon-disp}
\end{figure}

In Fig.~\ref{fig:ht-phon-disp} the phonon dispersion of \Co\ is shown
along selected lines of the first Brillouin-Zone (BZ) of the
orthorhombic HT phase. Along the path
W~($\frac{1}{2}, \frac{1}{2},
0$)--T~($\frac{1}{2}, \frac{1}{2}, \frac{1}{2}$) a low-frequency
branch with marginal dispersion can be identified. The displacement
pattern corresponding to the mode at W is in close analogy to the
static displacement pattern in the LT-phase of \Co.
\cite{Eickerling13} Furthermore, the path W--T of the low-lying phonon
branch can be correlated with the course of the diffuse rods in
reciprocal space (see Appendix \ref{appendix1}).\cite{Ishida73, Xu05}
Figuratively, its flat course can be interpreted in terms of equal
excitation energies for an infinite set of dynamic LT-phase-like
displacements of the cobalt and scandium atoms (illustrated in
Fig.~\ref{fig:struct-layers}b for a layer section in the $ab$-plane)
with differing modulations along the stacking direction
$c$.\footnote{The fact that the frequencies along the branch remain
  positive and the calculations do not predict the instability of the
  HT-phase most likely highlights the shortcomings of the standard GGA
  functional employed within this study to properly resolve the very
  flat energy surface region separating the HT- and LT-phase of \Co.}
The superposition of all these dynamic displacements yields the
picture of disorder between the layered building units of \Co. In
fact, similar behavior connected to weak coupling between layered
building units has been observed for other compounds such as the
francisite Cu$_3$Bi(SeO$_3$)$_2$O$_2$Cl. In the phonon dispersion of
its HT-phase, a nearly dispersionless branch connects a zone-center
mode at $\Gamma$ with equal atom displacements in all constituting
layers to a modulated variant of the mode at Z with layer-wise
inverted atom displacements.\cite{Milesi20}

To underline the correspondence between diffuse rods in X-ray
diffraction and a soft branch in the phonon dispersion we performed
simulations of the thermal diffuse scattering (TDS) contribution to
the diffracted intensity in the $(h, 1.5, l)$-plane. Consistent with
the inferences made above, simulations based on the phonon dispersion
of the HT-phase of \Co\ and assuming a temperature of 200~K
(Fig.~\ref{fig:ht-phon-disp}) reproduce the experimental observations
not only in the general positions and direction of the rods, but even
in details like the non-trivial intensity variations along the
$a^\ast$ direction. A comparison between the experimentally obtained
$(h, 1.5, l)$-plane at 200~K and the TDS simulation for the HT-phase
of \Co\ is given in Figs.~\ref{fig:tds-sim}a and c with corresponding
profiles in the top panel. The weak modulation of the simulated TDS
intensity with a period of two reciprocal lattice constants along
$c^\ast$ in Fig.~2c is most likely due to numerical artifacts
introduced by a Fourier interpolation step and not supported by
reciprocal space reconstructions based on X-ray diffraction
data.\footnote{Key properties of the phonon dispersion remain
  unaffected by this problem as is demonstrated in the Supplemental
  Material at [URL will be inserted by publisher].}

The phonon dispersion of the HT-phase of \Co, however, cannot explain
the gradual vanishing of the diffuse rods between 150~K and 80~K and
the appearance of pinpoint superstructure
reflections. Fig.~\ref{fig:tds-sim}b shows a reconstruction of the
$(h, 1.5, l)$-plane from measuring data at 12~K illustrating the
extensive reorganization of the X-ray diffraction pattern below
80~K. To account for these changes, a phonon softening mechanism may
be invoked that selectively reduces the frequency of the phonon mode
at W in the soft W--T branch to zero. Simultaneously with the
increasing dispersion between W and T comes a preference for
LT-phase-like dynamic atom displacements without superimposed
modulation along $c$ and a progressive structural ordering. Zero
phonon frequency at W is reached at 80~K resulting in the formation of
the LT-phase of \Co\ with static atom displacements from the
equilibrium positions in the HT-phase. Consequently, no more prominent
diffuse features are found in TDS simulations of the
$(h, 1.5, l)$-plane at 12~K employing the phonon eigenvalues of the
LT-phase\cite{Eickerling13} (see right part of
Fig.~\ref{fig:tds-sim}d).\footnote{A plot of the phonon dispersion for
  the LT-phase structure of \Co\ can be found in the Supplemental
  Material at [URL will be inserted by publisher].} Only localized TDS
contributions at positions of the experimentally observed
superstructure reflections remain, consistent with a doubling of the
unit cell. We note, that twinning was not considered within the
simulations and thus each second reflection position is missing in
Fig.~\ref{fig:tds-sim}d.  That the rearrangement of diffuse features
in reciprocal space cannot be attributed to temperature effects alone
is illustrated by TDS simulations for the HT-phase structure of \Co\
at 12~K (left part of Fig.~\ref{fig:tds-sim}d). Without a structural
transition weak diffuse rods would still be present at this
temperature. Again, parallels to the francisite
Cu$_3$Bi(SeO$_3$)$_2$O$_2$Cl may be drawn, where cooling induces a
successive frequency reduction of the phonon mode at Z in the soft
$\Gamma$-Z branch setting the stage to a displacive transition into a
static LT-phase with doubled $c$-parameter at
115~K.\cite{Constable17,Milesi20}

We may thus propose a consistent model for the observations in
electrical resistivity and intensity-weighted reciprocal space: In the
HT-regime (\textit{\textbf{I}}, see Fig.~\ref{fig:res-xrd}) above
150~K, dynamic disorder driven by the phonon modes W--T leads to the
occurrence of diffuse rods in reciprocal space. At approx. 150~K, the
mode at W starts to soften, thus continuously reducing the degree of
thermal fluctuations in the pre-LT regime (\textit{\textbf{II}}) and
leading to the successive ordering of the layers stacked in $c$
direction. Around 80~K, the softening process is complete and the
displacement pattern of the phonon mode freezes into the static atomic
positions observed in the LT-phase (\textit{\textbf{III}})
of \Co.

\section{Conclusion}\label{sec:conclusion}

To conclude, we provide experimental and theoretical evidence for a
soft-phonon-driven formation of a Peierls-type structurally distorted
state in \Co\ upon cooling. Based on the new results, the interplay
between two distinct transitions, \textit{i.e.} a charge-density wave
transition at 150~K and a Peierls-type distortion at 80~K, discussed in
earlier publications \cite{Scherer10,Scherer12,Eickerling13,Zhang12}
can now be consistently described by a single extended structural
phase transition \textit{via} an intermediate state between 80~K and
150~K. This intermediate state is characterized by phonon-driven
dynamic atom displacements in the crystallographic $ab$-plane with
strongly temperature-dependent frequency and correlation-length along
the $c$-axis. Inelastic X-ray or neutron scattering experiments might
provide further information on the progression of the phonon-softening
mechanism towards the static structurally distorted LT-phase observed
below 80~K.

\appendix
\section{Interpretation of diffuse X-ray scattering intensity --
  Relation to the phonon dispersion}\label{appendix1}
In the following, we outline the theoretical background for the
interpretation of the diffuse rods observed in reciprocal space for
\Co\ in temperature regimes \textbf{\textit{I}} and
\textbf{\textit{II}} (above $\approx$~80~K) in terms of Thermal
Diffuse Scattering (TDS).

We focus on the first-order TDS intensity
given by the sum over the absolute squares of the dynamical structure
factor contributions $F_j$ for all phonons $j$

\begin{equation}
  \label{eq:1}
  I_1(\mathbf{q}) = \frac{\hbar N I_e}{2} \sum_j{\frac{1}{\omega_{\mathbf{q},j}}
  \coth{\left(\frac{\hbar \omega_{\mathbf{q},j}}{2 k_B T}\right)} |F_j(\mathbf{q})|^2}
\end{equation}

\noindent where the number of unit cells is denoted by $N$ and the one-electron scattering
intensity by $I_e$. Every sum contribution is weighted by a thermal
occupation factor $\frac{1}{\omega_{\mathbf{q},j}}
\coth{\left(\frac{\hbar \omega_{\mathbf{q},j}}{2 k_B T}\right)}$ for phonon
$j$ at wave vector $\mathbf{q}$ with frequency
$\omega_{\mathbf{q},j}$.\cite{Xu05} Thus, the contributions of low-frequency
phonons to TDS are dominant. The dynamical structure factors $F_j$ derive
from the phonon-specific shift of the atoms $s$ in the unit cell as

\begin{equation}
  \label{eq:2}
  F_j(\mathbf{q}) = \sum_s{\frac{f_s}{\sqrt{\mu_s}} \exp{(-M_s)}
  \left(\mathbf{q} \cdot \mathbf{e}_{\mathbf{q},j,s}\right)}
\end{equation}

\noindent with atomic scattering factor $f_s$, atomic mass $\mu_s$ and
Debye-Waller factor $M_s$. Size and direction of the atom shifts are
described by the polarization vector $\mathbf{e}_{\mathbf{q},j,s}$.
\cite{Xu05}  Special attention should be paid to the double meaning
of the wave vector $\mathbf{q}$: On the one hand, $\mathbf{q}$ is the
sum of the reduced phonon wave vector in the first Brillouin zone
$\mathbf{k}$ and a reciprocal lattice vector $\mathbf{K}_\mathbf{q}$

\begin{equation}
  \label{eq:3}
  \mathbf{q} = \mathbf{k} + \mathbf{K}_\mathbf{q}
\end{equation}

\noindent Due to the periodicity of the phonon dispersion in reciprocal space,
$\omega_{\mathbf{q},j}$ and $\omega_{\mathbf{k},j}$ are identical. On the other
hand, $\mathbf{q}$ indicates the position of the considered TDS
feature and $\mathbf{K}_\mathbf{q}$ the position of the adjacent
Bragg reflection in reciprocal space.

The dominant diffuse reciprocal space features in the HT regime of \Co\
are the rods along $c^\ast$. The location of
all points on the diffuse rods with respect to the adjacent
Bragg reflections can be described by wave vectors
$\mathbf{k} = \left(\frac{1}{2}, \frac{1}{2}, l\right)$ with $l$
varying between 0 and $\frac{1}{2}$. Thereby, the boundary points
$\mathbf{k} = \left(\frac{1}{2}, \frac{1}{2}, 0\right)$ and
$\mathbf{k} = \left(\frac{1}{2}, \frac{1}{2}, \frac{1}{2}\right)$
represent positions on and exactly midway between the superstructure
reflections appearing in the pre-LT and LT regime. From this
information, a low-frequency, flat phonon branch connecting the
zone-boundaries at W ($\hat{=}~\frac{1}{2}, \frac{1}{2}, 0)$ and
T ($\hat{=}~\frac{1}{2}, \frac{1}{2}, \frac{1}{2}$) may
be postulated for the HT regime. Indeed, an acoustic phonon mode with extraordinarily
low frequency between W and T is
found in the ab-initio phonon dispersion relation based on the
idealized ground-state structure of HT \Co\ (highlighted in
green in Fig.~\ref{fig:ht-phon-disp}). 

\section{Interpretation of diffuse X-ray scattering intensity -- Intensity modulations}\label{appendix2}
\begin{figure}
  \includegraphics[width=0.48\textwidth]{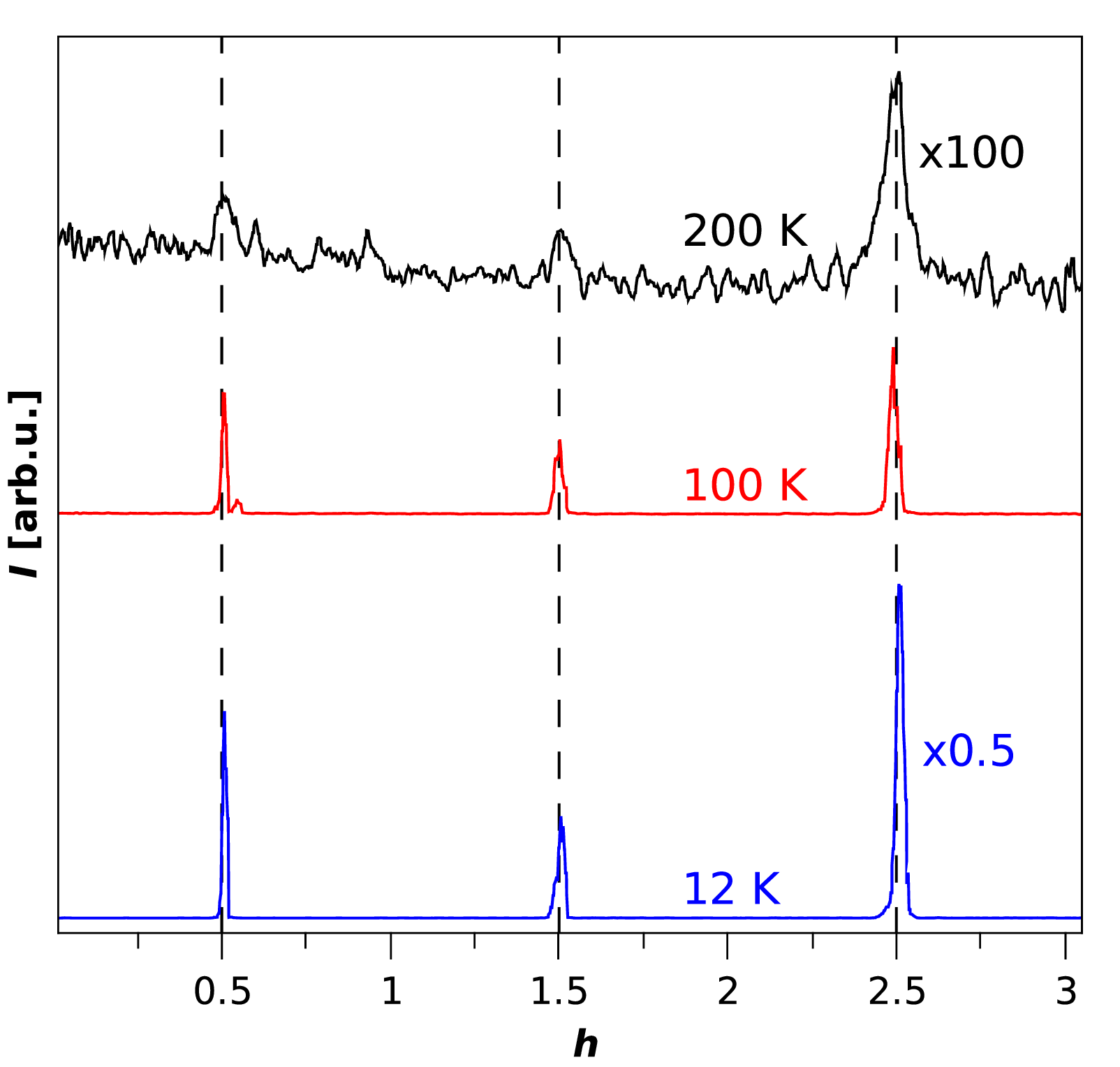}
  \caption{\label{fig:int-pattern-superstructure-diff} Intensity
    profiles along $[h, 1.5, 0]$ at 200~K (HT regime), at 100~K
    (pre-LT regime) and at 12~K (LT regime). All reflection indices
    refer to the orthorhombic HT-phase unit cell of \Co. For clarity,
    the profiles at 200~K and 12~K were scaled by factors of 100 and
    0.5, respectively. In all cases, an alternation between
    strong and weak intensity along $h$ becomes evident.} 
\end{figure}

Focusing on the intensity modulation of the TDS, diffuse rods and
pinpoint superstructure reflections show similar intensity
alternations along the reciprocal space directions $a^\ast$ and
$b^\ast$ (see Fig.~\ref{fig:tds-sim}).  More specifically, rows of
weaker superstructure reflections or diffuse rods along $c^\ast$ are
invariably followed by rows of stronger superstructure reflections or
diffuse rods and vice versa.  Thereby, the intensity of both, weak and
strong reciprocal space features, rises with increasing $h$- or
$k$-index. In Fig.~\ref{fig:int-pattern-superstructure-diff},
representative line profiles of the scattering intensity along
$[h, 1.5, 0]$ cutting through the superstructure reflection positions
in the $(h, 1.5, l)$-plane are shown for each of the temperature
regimes of \Co\ postulated in section~\ref{sec:discussion}. The
similarity of the line profiles in
Fig.~\ref{fig:int-pattern-superstructure-diff} may be taken as a hint
to \textit{(i)}~the common origin of diffuse rods and pinpoint
superstructure reflections and \textit{(ii)}~a persistent displacive
modulation of the atom positions in the $ab$-plane with
temperature-dependent correlation length along $c$ as opposed to
simple stacking disorder.

In the following, we will relate the characteristic scattering
intensity variation along the line $[h, 1.5, 0]$ to a simplified model
of the displacive modulation pattern in \Co.  Only scattering
contributions from the set of atoms most affected by shifts from their
high-symmetry positions, {\it i.e.} the cobalt and the scandium atoms capping
the $\left[\right.$Co(C$_2$)$_2$Co$\left.\right]$ hexagons (see Fig.~\ref{fig:struct-layers}),
are taken into account. The effect of the shifts in the carbon atom positions is
neglected due to the relatively small scattering cross-section of
carbon compared to scandium and cobalt. Furthermore, only the
scattering contributions from one of two twin domains are considered
(\textit{vide supra}). The contributions of the second domain may be
symmetry-generated by application of the according twin law.

In a first step, the contributions of the cobalt and
scandium atoms to the structure factor $F(\mathbf{h})$ are treated
separately. The general structure factor equation for a one-atom
structure subjected to an arbitrary displacement wave is detailed
elsewhere.\cite{neder-f-displac-wave,Giacovazzo_displac_wave}
Assuming a perfectly commensurate
modulation the structure factor of \Co\ at
the position of superstructure reflections can be described by
the simplified formula

\begin{eqnarray}
  \label{eq:f_displac_mod}
  F(\mathbf{h}) &=& J_1(2 \pi \mathbf{ha})~\exp(i
  \phi)~G\left(\mathbf{h} + \mathbf{q}\right) \\*\nonumber
  &-&
  J_1(2 \pi \mathbf{ha})~\exp(-i \phi)~G\left(\mathbf{h} - \mathbf{q}\right)
\end{eqnarray}

\noindent The position of the superstructure reflections is referred to
by the reciprocal space vector $\mathbf{h}$, while $J_1(2 \pi \mathbf{ha})$
and $G(\mathbf{h})$ denote the structure factor of the unmodulated
structure and the first-order Bessel function, respectively. The displacement
wave is characterized by its amplitude vector $\mathbf{a}$, wave vector
$\mathbf{q}$ and phase $\phi$.

\begin{figure}
  \centering
  \includegraphics[width=0.48\textwidth]{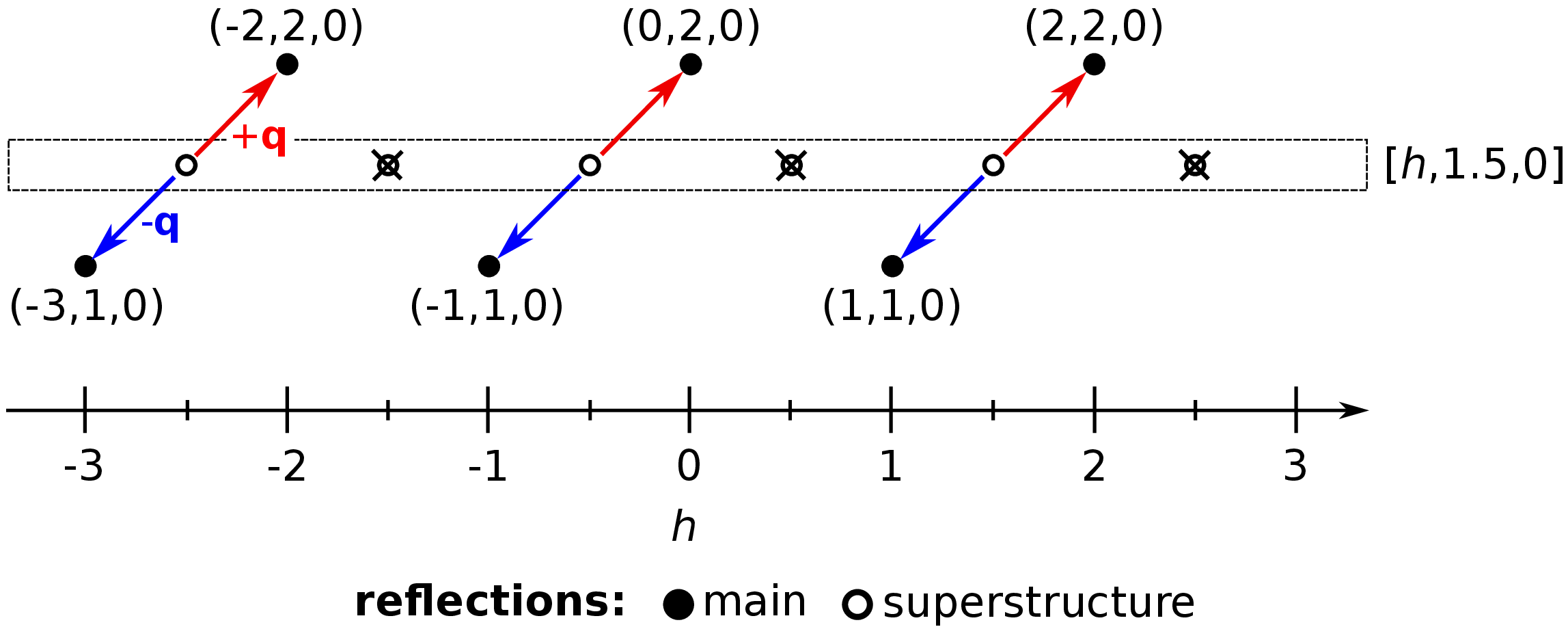}
  \caption{Position of the main reflections in the $(h, k, 0)$-plane
    contributing to the structure factor of the superstructure
    reflections along the line $[h, 1.5, 0]$. All reflection indices
    are referred to the orthorhombic HT-phase unit cell of \Co. The wave vectors
    $+\mathbf{q}$ and $-\mathbf{q}$ connecting the superstructure
    reflections to the main reflections are displayed in red and
    blue, respectively. Absent superstructure reflections are
    crossed out (see text).}
  \label{fig:main_superstruc_refl_pos}
\end{figure}

Concerning the cobalt substructure, a phase $\phi = 0$ may be put
for the displacements\footnote{Only the relative
phase of the cobalt and scandium atom displacements is of interest for the
final value of $F(\mathbf{h})$. Thus, the value of $\phi$ for either
the cobalt or the scandium atom displacements may be chosen arbitrarily as 0 or
$\pi$.} leading to 

\begin{eqnarray}
  \label{eq:f_co_displac}
  F_{\mathrm{Co}}(\mathbf{h}) &=&
                                  J_1(2 \pi \mathbf{ha}) \cdot
                                  G_{\mathrm{Co}}\left(\mathbf{h} + \mathbf{q}\right)\\*\nonumber
                              &-&
                                  J_1(2 \pi \mathbf{ha}) \cdot 
                                  G_{\mathrm{Co}}\left(\mathbf{h} - \mathbf{q}\right)
\end{eqnarray}

\noindent so that in a first step the variation of
$J_1(2 \pi \mathbf{ha})$ and $G(\mathbf{h})$ along the line
$[h, 1.5, 0]$ must be considered. As the cobalt atom displacements are directed
along the crystallographic $a$-direction ($\mathbf{a} = (a, 0, 0)^T$),
the argument of $J_1$ reduces to the scalar $ha$.  Non-zero values of
$G_{\mathrm{Co}}\left(\mathbf{h} \pm \mathbf{q}\right)$ may only
occur, if a vector $\pm \mathbf{q} = \pm (0.5, 0.5, 0)^T$ connects the
reciprocal space position $\mathbf{h}$ of the considered
superstructure reflection to the position of a non-extinct main
reflection.\footnote{A wave vector $\mathbf{q} = (0.5, 0.5, p)^T$ must
  be used in the calculation of the superstructure reflection
  intensity along arbitrary lines $[h, 1.5, p]$ parallel to
  $[h~1.5~0]$.}  In accord with the two possible signs of
$\mathbf{q}$, there are always two contributing main reflection
positions in the $(h, k, 0)$-plane for each superstructure reflection (see
Fig.~\ref{fig:main_superstruc_refl_pos}).

\begin{figure}
  \includegraphics[width=0.48\textwidth]{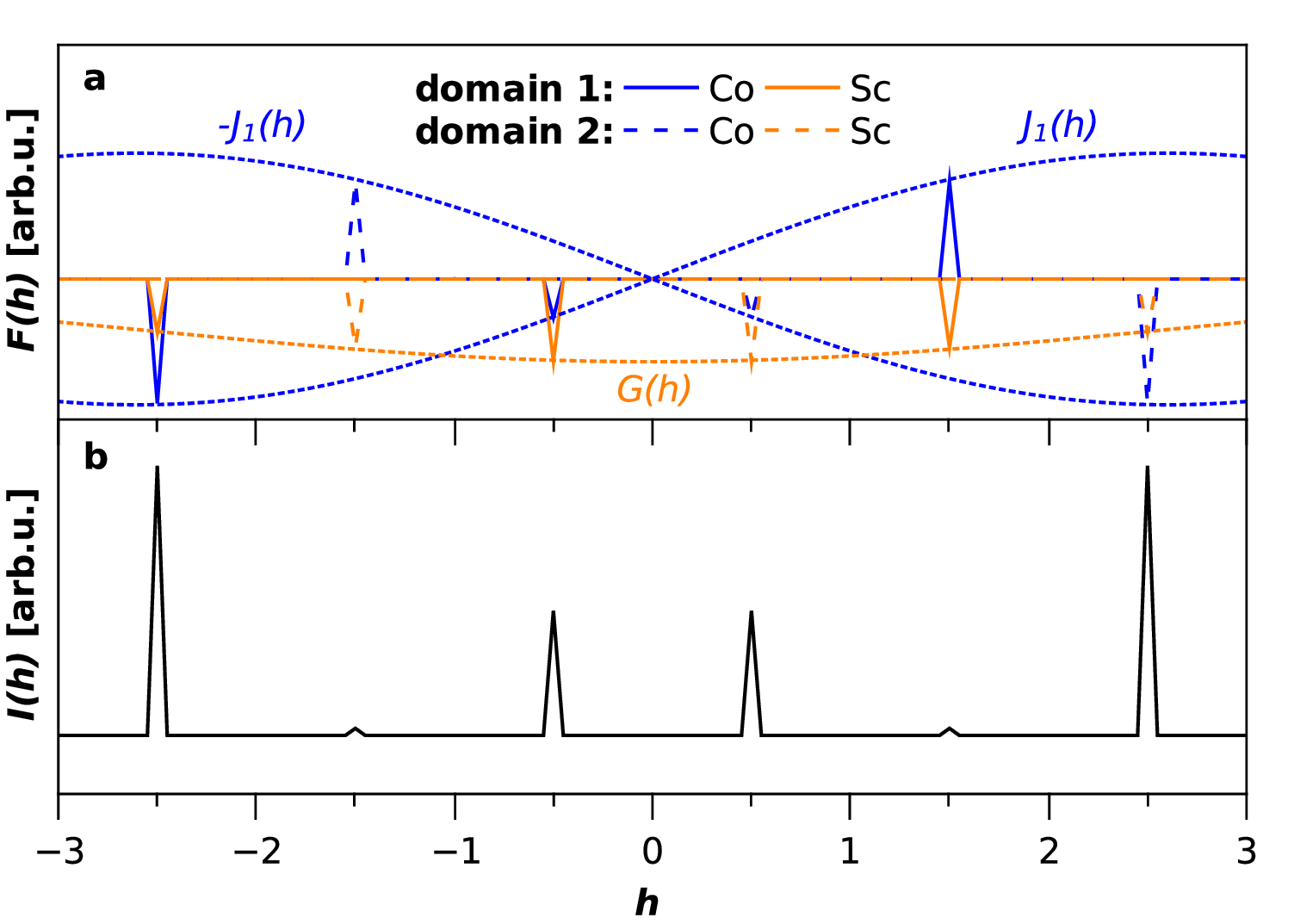}
  \caption{\label{fig:struc-fac-contribs} Simulation \cite{Proffen97} of the contributions to
    $F(\mathbf{h})$ along $[h, 1.5, 0]$ (referring to the
    orthorhombic HT-phase unit cell of \Co) for a simplified modulated \Co\
    structure~(a): cobalt atom contributions are
    given in blue with tentative enveloping first-order Bessel
    functions $J_1(h)$ and $-J_1(h)$.  Scandium atom contributions are given in
    red with an enveloping Gaussian function $G(h)$ as approximation
    for the angle-dependence of the atomic scattering factor of
    scandium. Scattering due to domain~1 is signified by solid lines,
    scattering due to domain~2 by hatched lines. The summed up
    and squared structure factor contributions (b) reproduce the
    experimentally observed intensity alternation scheme
    along $[h, 1.5, 0]$.}
\end{figure}

However, every second
superstructure reflection along $[h, 1.5, 0]$ is absent (crossed out in
Fig.~\ref{fig:main_superstruc_refl_pos}), because its associated main
reflections are extinct due to the body-centering of the HT \Co\
crystal structure. Due to the small variation of the atomic scattering
factors over the narrow considered range of $h$-values the structure
factors $G_{\mathrm{Co}}\left(\mathbf{h} \pm \mathbf{q}\right)$ for
the contributing main reflections of the unmodulated Co
substructure\footnote{In the present simplified structure containing
  only cobalt atoms, two cobalt atoms at the fractional coordinates
  $(0, 0.5, 0)$ and $(0.5, 0, 0.5)$ have to be taken into account in
  the calculation of
  $G_{\mathrm{Co}}\left(\mathbf{h} \pm \mathbf{q}\right)$.}  depend
only weakly on $h$ with
$G_{\mathrm{Co}}\left(\mathbf{h} + \mathbf{q}\right) > 0$ and
$G_{\mathrm{Co}}\left(\mathbf{h} - \mathbf{q}\right) \approx
-G_{\mathrm{Co}}\left(\mathbf{h} + \mathbf{q}\right)$.  Thus, the
variation of the superstructure reflection structure factor
$F_{\mathrm{Co}}(\mathbf{h})$ along $[h, 1.5, 0]$ in good approximation
mirrors the behavior of the Bessel function $J_1(2 \pi ha)$ (see solid
blue line in Fig.~\ref{fig:struc-fac-contribs}a).

The same strategy can be applied in the derivation of
the structure factor contribution of the scandium atoms $F_{\mathrm{Sc}}(\mathbf{h})$.
Displacements of the scandium atoms towards
long Co$\cdots$Co contacts\footnote{This is the situation found in the
  LT structure of \Co\ published earlier, see Ref. \citenum{Eickerling13}.} are accounted
for by putting a value of $\pi$ for the displacement wave phase $\phi$, so
that 

\begin{eqnarray}
  \label{eq:f_sc1_displac}
  F_{\mathrm{Sc}}(\mathbf{h}) =
  &-&J_1(2 \pi \mathbf{ha}) \cdot
                                  G_{\mathrm{Sc}}\left(\mathbf{h} + \mathbf{q}\right) \\*\nonumber
                              &+&
  J_1(2 \pi \mathbf{ha}) \cdot
  G_{\mathrm{Sc}}\left(\mathbf{h} - \mathbf{q}\right)
\end{eqnarray}

\noindent is obtained. In contrast to the cobalt atoms, the scandium
atom displacements point along the crystallographic $b$-direction
($\mathbf{a} = (0, a, 0)^T$), so that the argument $\mathbf{ha}$ of
the Bessel functions in Eq.~\ref{eq:f_sc1_displac} is constant along
$[h, 1.5, 0]$ and $J_1(\mathbf{ha}) = const. > 0$ over the whole range
of $h$-values. Approximating the structure factor for the contributing
main reflections of the unmodulated scandium
substructure\footnote{Only the strongly displaced scandium atoms at
  the fractional coordinates $(0.5, 0, 0)$ and $(0, 0.5, 0.5)$ are
  taken into account.}  in analogy to the cobalt atom substructure
again yields a weakly $h$-dependent value of
$G_{\mathrm{Sc}}\left(\mathbf{h} \pm \mathbf{q}\right)$ with
$G_{\mathrm{Sc}}\left(\mathbf{h} + \mathbf{q}\right) > 0$ and
$G_{\mathrm{Sc}}\left(\mathbf{h} - \mathbf{q}\right) \approx
-G_{\mathrm{Sc}}\left(\mathbf{h} + \mathbf{q}\right)$.  Therefore, the
structure factor contribution $F_{\mathrm{Sc}}(\mathbf{h})$ of the Sc
atoms at the superstructure reflection positions along $[h, 1.5, 0]$
shows only slight variation with $h$ and is invariably negative (see
solid red line in Fig.~\ref{fig:struc-fac-contribs}a).  Finally, the
structure factor contributions $F_{\mathrm{Co}}(\mathbf{h})$ and
$F_{\mathrm{Sc}}(\mathbf{h})$ of cobalt and scandium along
$[h, 1.5, 0]$ can be summed up and also the structure factor
contributions of the second \Co\ domain can be generated at this stage
by mirroring the contributions from the first domain at the
$y$-axis\footnote{More precisely, a twofold rotation of the
  diffraction pattern about $a^\ast$ has to be performed corresponding
  to a twofold rotation about $a$ as twin law in real-space.} (hatched
lines in Fig.~\ref{fig:struc-fac-contribs}a).  At every second
superstructure reflection position (counted from $h = 0$) opposite
signs of $F_{\mathrm{Sc}}(\mathbf{h})$ and
$F_{\mathrm{Co}}(\mathbf{h})$ meet to yield a small absolute value of
the total structure factor $F(\mathbf{h})$. In contrast, at
superstructure reflection positions with uneven number equal signs of
$F_{\mathrm{Sc}}(\mathbf{h})$ and $F_{\mathrm{Co}}(\mathbf{h})$ meet
to yield a large absolute value of $F(\mathbf{h})$. Taking the
absolute square of the summed structure factor profile leads to the
observed characteristic intensity modulation for diffuse rods and
pinpoint superstructure reflections along $[h, 1.5, 0]$ in \Co\ (see
Fig.~\ref{fig:struc-fac-contribs}b).

\end{document}



\title{Supporting Information for: Evidence for a soft-phonon mode driven
  Peierls-type distortion in \Co}


\author{Jan Langmann}
\author{Christof Haas}
\affiliation{University of Augsburg, Institut f\"ur Physik,
  Universit\"at Augsburg,
  Universit\"atsstra\ss e 1, D-86159 Augsburg, Germany}
\author{Emmanuel  Wenger}
\author{Dominik Schaniel}
\affiliation{Université de Lorraine, CNRS, CRM2, F-54000 Nancy, France}
\author{Wolfgang Scherer}
\author{Georg Eickerling}
\email{georg.eickerling@physik.uni-augsburg.de}
\affiliation{University of Augsburg, Institut f\"ur Physik,
  Universit\"at Augsburg,
  Universit\"atsstra\ss e 1, D-86159 Augsburg, Germany}


\date{\today}


\pacs{}

\maketitle


\tableofcontents

\newpage

\section{Investigated samples}
\label{sec:investigated-samples}

Two types of \Co\ single-crystals were investigated in this work:
Needle-shaped single-crystals obtained according to methods described
in Refs.~\citenum{Rohrmoser07,Vogt09,He15} and platelet-like
single-crystals from heat treatment of \Co\ powder in a lithium
flux.\cite{Haas19}

\begin{figure}[p]
  \includegraphics[width=1.0\textwidth]{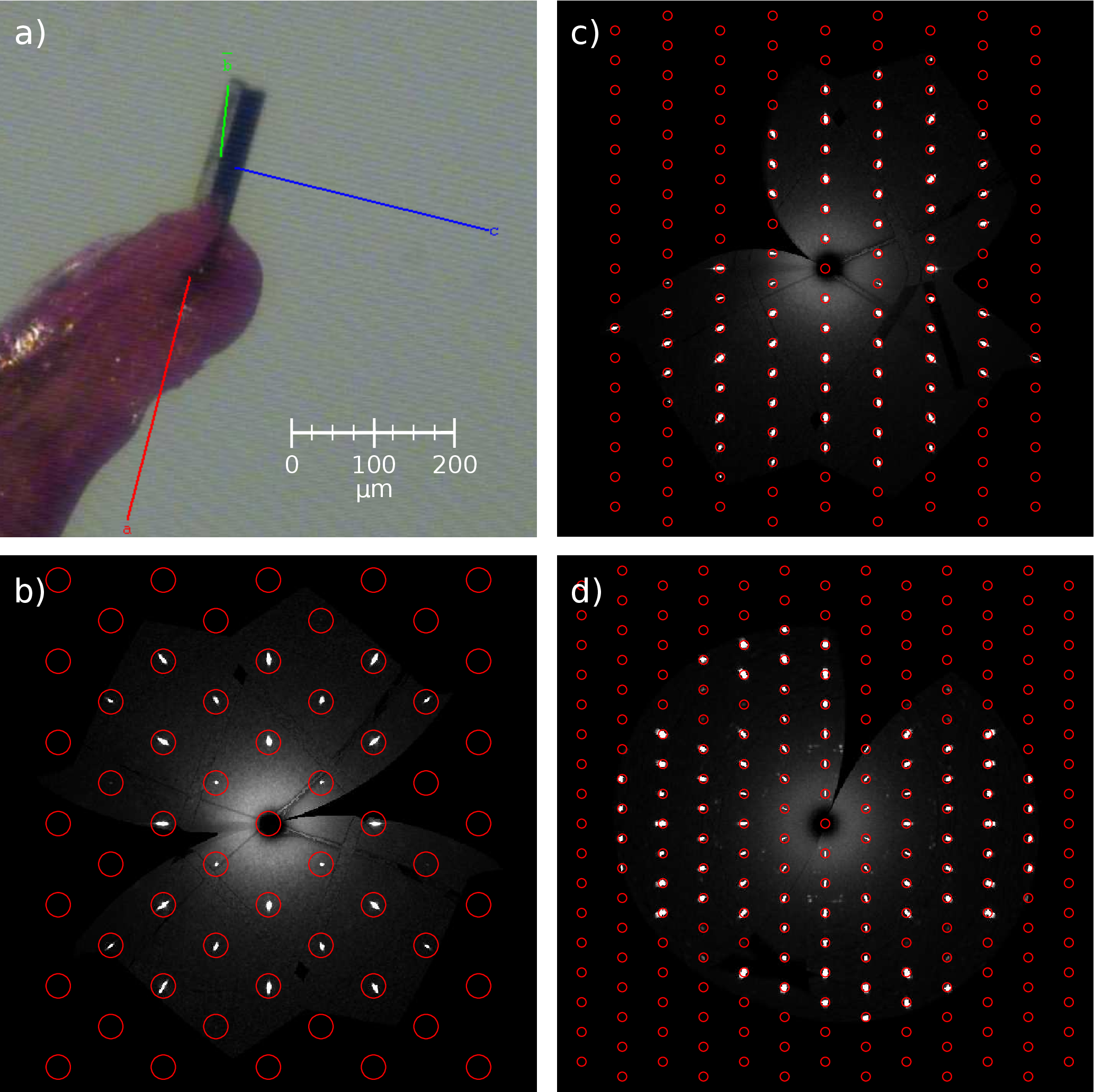}
  \caption{(a)~Photographic image of a typical needle-shaped \Co\
    single-crystal with crystal axes $a$, $b$ and $c$ referring to the
    orthorhombic high-temperature phase unit cell indicated by
    coloured lines. Reconstructions of reciprocal-space
    planes (b)~$(hk0)$, (c)~$(h0l)$ and (d)~$(0kl)$ from
    room-temperature X-ray diffraction data. Predicted reflection
    positions for HT-\Co\ are indicated by red circles.}
  \label{fig:needle}
\end{figure}

\begin{figure}[p]
  \centering
  \includegraphics[width=0.85\textwidth]{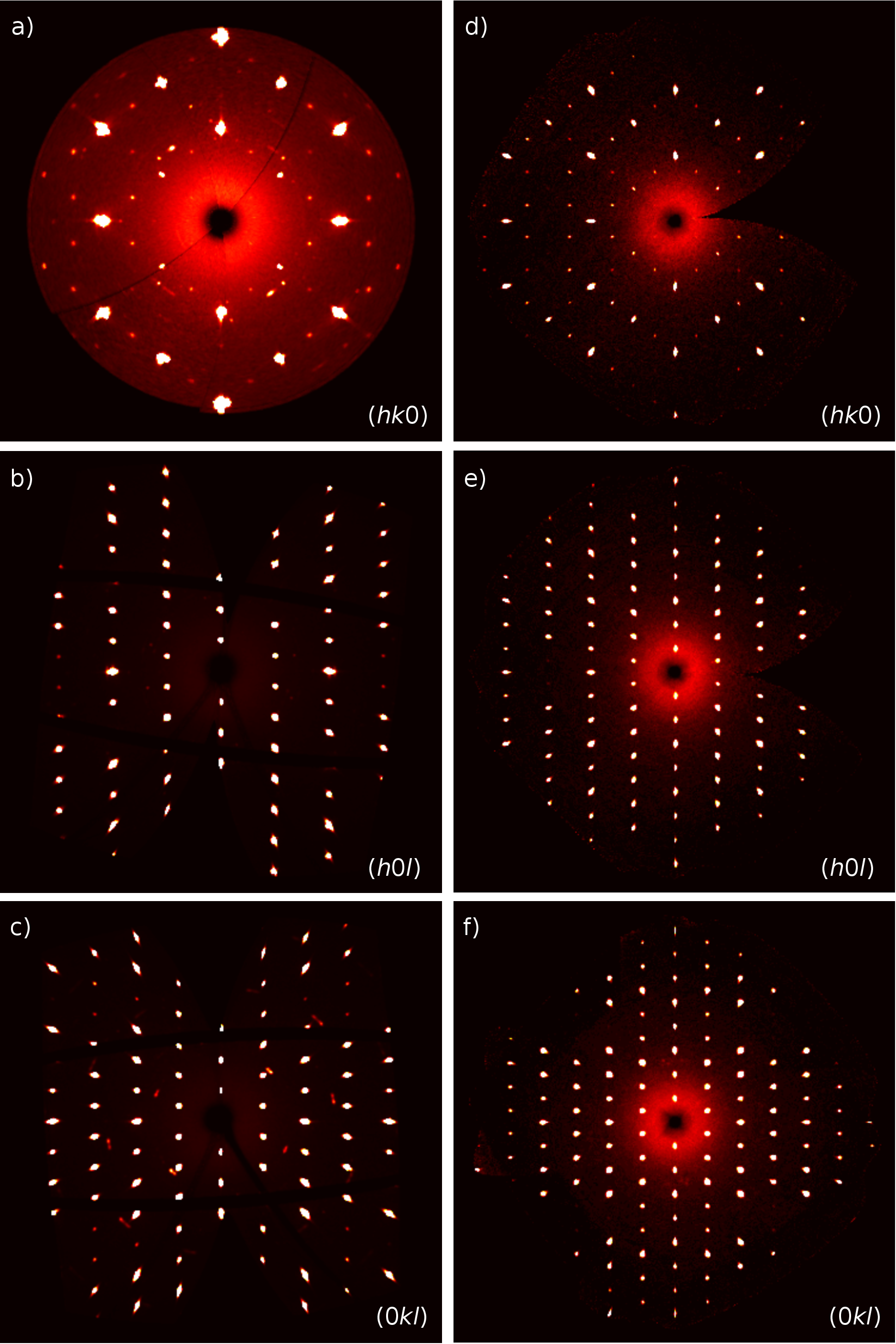}
  \caption{Reconstructions of reciprocal-space planes $(hk0)$, $(h0l)$
    and $(0kl)$ from X-ray diffraction data for a platelet-like
    (a-c) and a needle-shaped \Co\ single-crystal (d-f) at 150~K.}
  \label{fig:diffraction-data-150K}
\end{figure}

The needle-shaped single-crystals were used in the determination of
the temperature-dependent X-ray scattering intensity of pinpoint
superstructure reflections and the tempera\-ture-dependent electrical
resistivity.  Sample sizes were characterized by an approximate
thickness of 20~$\mu$m and an approximate length of 200~$\mu$m (see
Fig.~\ref{fig:needle}a for a phonographic image of a typical
sample). Reconstructions of common reciprocal-space planes from
room-temperature X-ray diffraction data (Figs.~\ref{fig:needle}b-d)
indicate high crystalline quality with only minor imperfections.
Comparison with corresponding reciprocal-space planes at 150~K
(Figs.~\ref{fig:diffraction-data-150K}a-c) underlines that the sample
quality is not degraded by systematic twinning in the transition from
the high-temperature to the low-temperature phase of \Co.

\begin{figure}[p]
  \centering
  \includegraphics[width=1.0\textwidth]{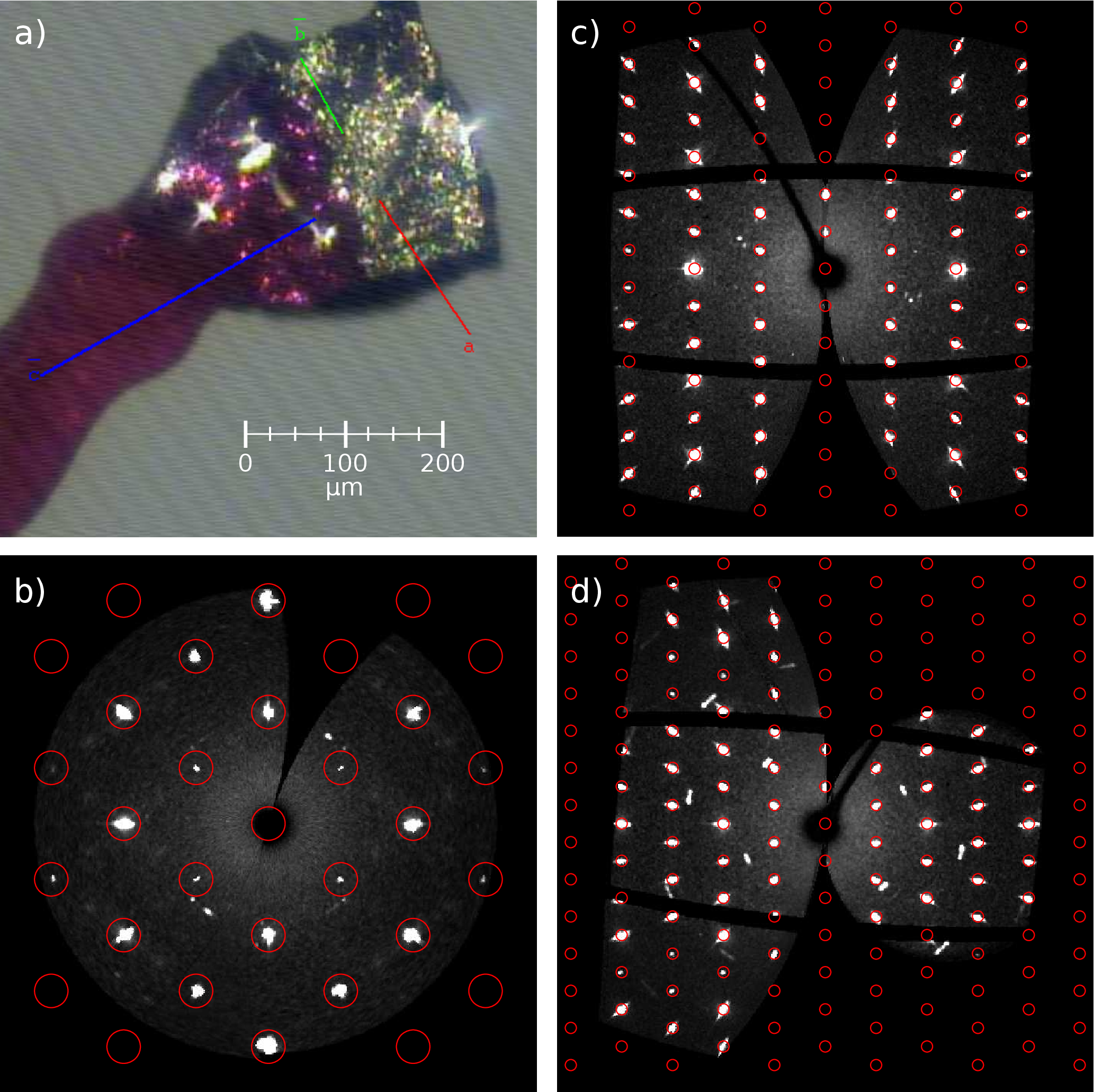}
  \caption{(a)~Photographic image of a platelet-like \Co\
    single-crystal with crystal axes $a$, $b$ and $c$ referring to the
    orthorhombic high-temperature phase unit cell indicated by
    coloured lines. Reconstructions of reciprocal-space planes
    (b)~$(hk0)$, (c)~$(h0l)$ and (d)~$(0kl)$ from room-temperature
    X-ray diffraction data. Predicted reflection positions are
    indicated by red circles.}
  \label{fig:platelet}
\end{figure}

Due to their larger size with an approximate thickness of 150~$\mu$m
and an approximate lateral extension of 300~$\mu$m (see
Fig.~\ref{fig:platelet}a for a photographic image of the sample) the
platelet-like single-crystals were employed in the
temperature-dependent tracking of diffuse X-ray scattering
features. Significant, but non-interfering crystal imperfections are
apparent from scattered intensity at non-indexed positions in
reconstructed reciprocal-space planes at room-temperature
(Figs.~\ref{fig:platelet}b-d) and 150~K
(Figs.~\ref{fig:diffraction-data-150K}d-f).

\begin{figure}[h]
  \centering
  \includegraphics[width=0.5\textwidth]{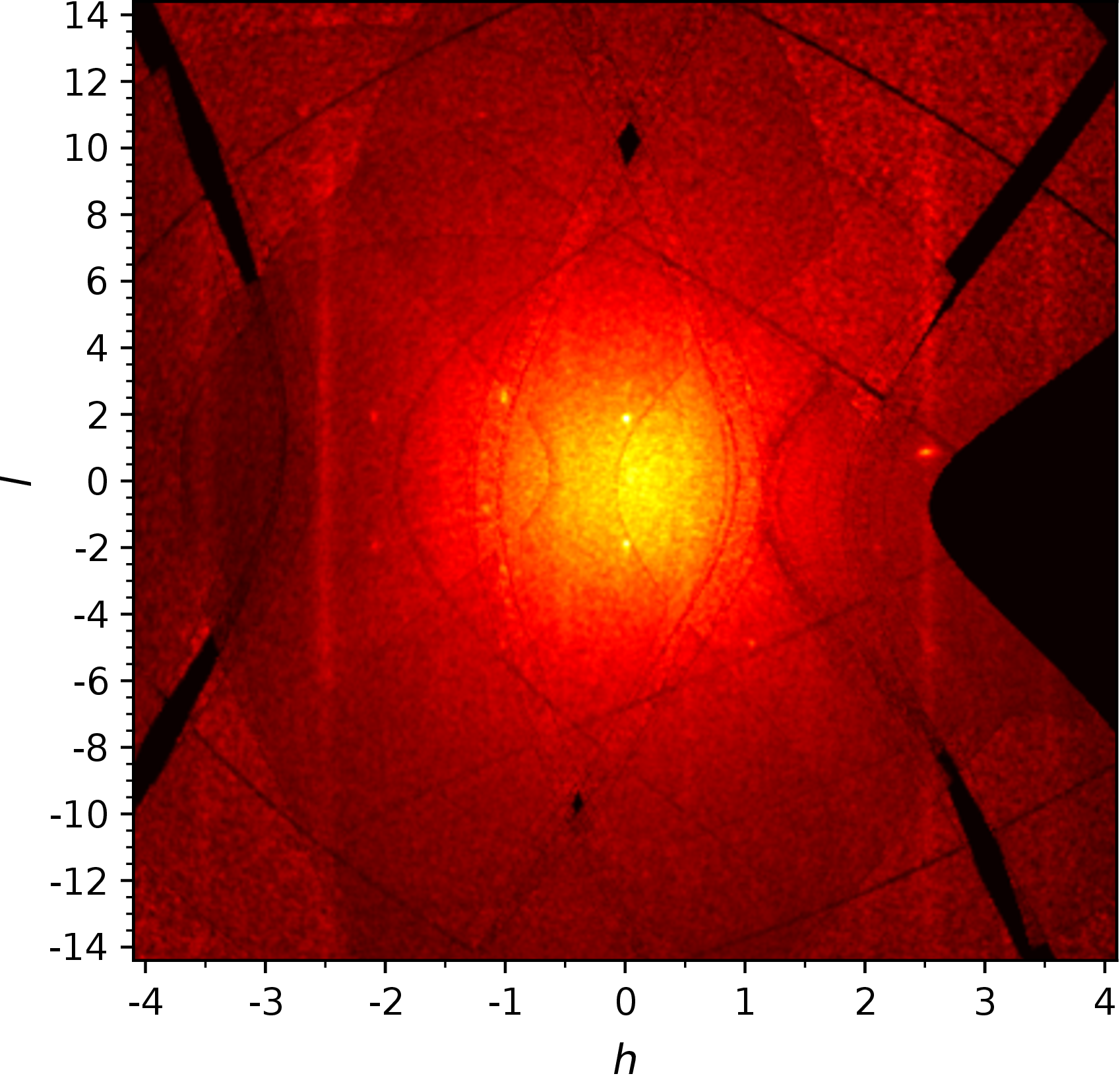}
  \caption{Diffuse X-ray scattering features in the $(h,1.5,l)$-plane
    for a needle-shaped single-crystal at room-temperature.}
  \label{fig:needle-diffuse}
\end{figure}

\begin{figure}[h]
  \centering
  \includegraphics[width=0.57\textwidth]{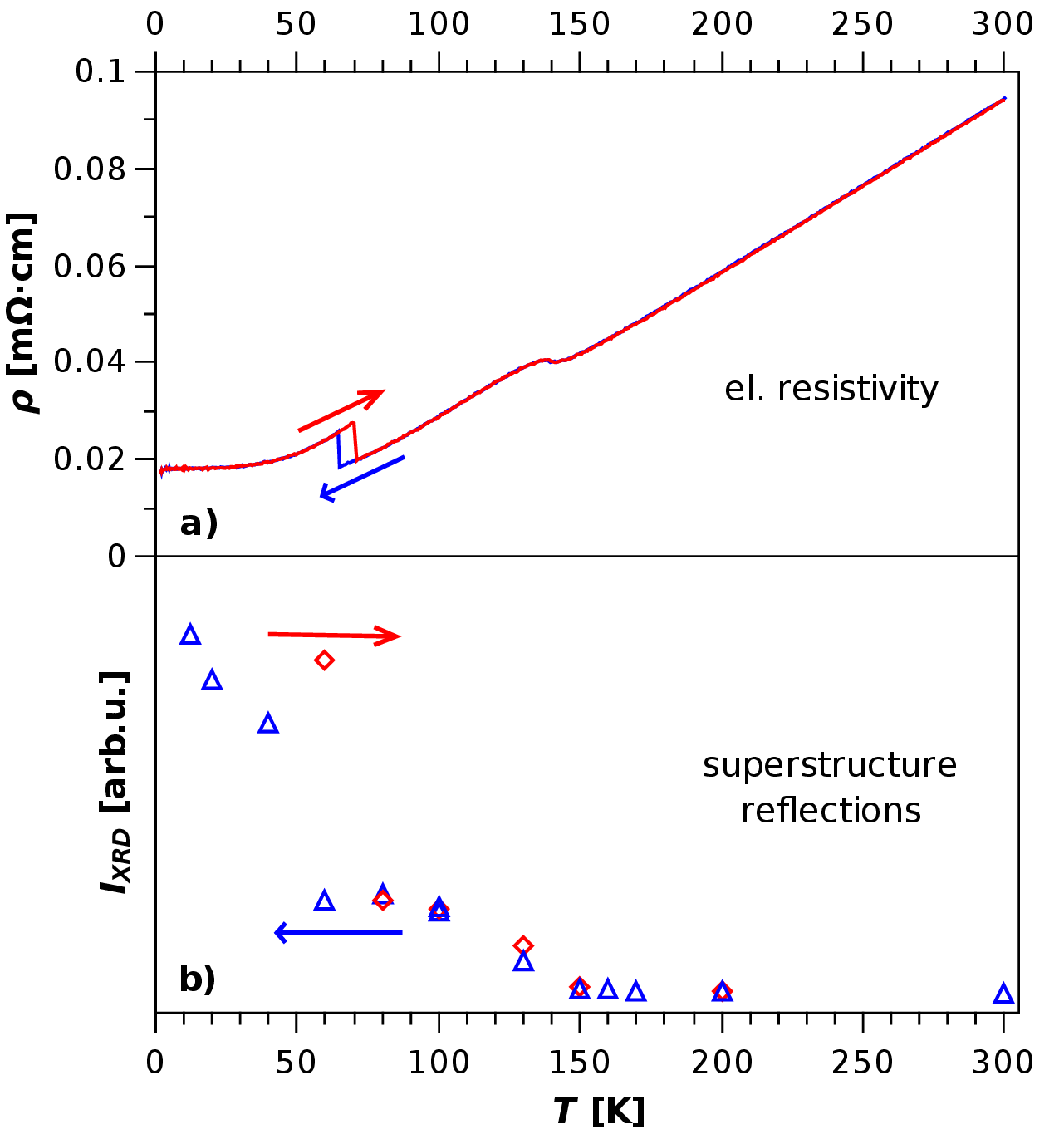}
  \caption{Temperature-dependence of (a)~the electrical resistivity
    \rhoT\ and (b)~the X-ray scattering intensity at superstructure
    reflection positions \issrT\ for a platelet-like single-crystal.}
  \label{fig:platelet-ixrd-rho}
\end{figure}

In order to verify, that the type of the sample does not affect the
results discussed in the main text, measurements were performed on
both types of single-crystals. In analogy to the platelet-like
single-crystals, weak diffuse rods can be recognized in the
room-temperature reconstruction of the $(h,1.5,l)$-plane for a
needle-shaped single-crystal
(Fig.~\ref{fig:needle-diffuse}). Complementarily, the characteristic
two-step increase of the superstructure reflection intensity \issrT\
found for the needle-shaped single-crystals is reproduced by a
platelet-like single-crystal (Fig.~\ref{fig:platelet-ixrd-rho}a). This
also holds for the corresponding two jumps in the electrical
resistivity \rhoT\ (Fig.~\ref{fig:platelet-ixrd-rho}b). Sample size
effects may account for the slight downward shift of the anomalies in
\issrT\ and \rhoT\ as compared to the needle-shaped
single-crystals.\cite{Natarajan69} Notably, the abrupt nature of the
lower anomaly in scattering intensity and electrical resistivity is
further emphasized by a sizable temperature hysteresis between cooling
and heating cycle for the platelet-like single-crystal.

\FloatBarrier

\newpage

\section{X-ray diffraction experiments}\label{sec:x-ray}

The temperature-dependence of the superstructure reflection intensity
was determined by means of variable-temperature X-ray diffraction
experiments on a single-crystalline needle of \Co\ mounted on a capton
mesh (MiTeGen). Depending on the temperature range two different
experimental setups were employed: Sample temperatures in the range
100~K~$<T<$~300~K were reached on a CAD4 $\kappa$-goniometer (BRUKER)
fitted with a micro-focus tube (INCOATEC, $\lambda$(Mo~K$_\alpha$) =
0.71073\AA) and an open-flow N$_2$ gas stream cooler
(OXFORD).\cite{Cosier86} The scattering intensity was recorded in
$\omega$-scans ($\Delta\omega$ = 2\degr, $t$ = 582.5~s) on an XPAD~3.2
hybrid pixel detector.\cite{Wenger14} $\phi$-scans ($\Delta\phi$ =
0.5\degr, $t$ = 30~min) in the temperature range 10~K$<T<$100~K were
performed employing a Huber Eulerian cradle goniometer equipped with a
modified closed-cycle He cryostat (ARS Cryo),\cite{Reisinger07} a
MAR345 image plate detector (MARXPERTS), and a graphite monochromated
FR591 rotating anode with molybdenum target (BRUKER,
$\lambda$(Mo~K$_\alpha$) = 0.71073\AA).

An updated version of the latter setup featuring a micro-focus tube
(INCOATEC, $\lambda$(Ag~K$_\alpha$)~= 0.56087~\AA) and a Pilatus3 R
CdTe detector (DECTRIS) was used to collect long-exposure diffraction
data for a lithium flux grown plate shaped single crystal of \Co\
(12~K~$< T <~$100~K, $\phi$-Scans, $\Delta \phi$ = 0.5\degr, $t$ =
120~s). The sample was glued to a capton microloop (MiTeGen) using nail varnish.
In the temperature range 100~K~$< T <$~300~K the closed-cycle He
cryostat was replaced by an open-flow N$_2$ gas stream cooler
(OXFORD).\cite{Cosier86}

\newpage

\section{Twinning}\label{sec:twinning}

A direct observation of the twin domains in \Co, e.g. by means of
transmission electron microscopy (TEM), has not been achieved so
far. Yet, inferences about the most probable orientation of the twin
domain boundaries can be made on group theoretical grounds. Hints to
the domain sizes come from single-crystal X-ray diffraction data.

As already pointed out by Vogt \textit{et al.}\cite{Vogt09},
systematic twinning in \Co\ is induced by a $t2$ step in the symmetry
reduction pathway from the high-temperature to the low-temperature
phase space-group $I\frac{2}{m}\frac{2}{m}\frac{2}{m}$
$\overset{t2}{\rightarrow}$ $I11\frac{2}{m}$
$\overset{i2}{\rightarrow}$ $B11\frac{2}{m}$
($\hat{=}$~$C1\frac{2}{m}1$). Thereby, twin domains with two distinct
orientation states are formed. Consideration of the point groups of
the involved space groups is sufficient for a determination of the
twin operations relating the two possible twin domain orientations to
each other.\cite{ITC-VolD-Twin} Coset decomposition of the point group
$\frac{2}{m}\frac{2}{m}\frac{2}{m}$
($I\frac{2}{m}\frac{2}{m}\frac{2}{m}$) with respect to the point group
$11\frac{2}{m}$ ($I11\frac{2}{m}$) yields the set of formally
equivalent twin operations
$\{m_{[100]}, 2_{[100]}, m_{[010]}, 2_{[010]}\}$. The mirror plane
$m_{[100]}$ has the highest chance of actually defining the
macroscopic twin element that partitions the crystal into twin
domains. This comes for two reasons: In the formation of twin domains
mirror planes mostly take precedence over rotation
axes.\cite{ITC-VolD-Twin} Additionally, hypothetical twin domain
boundaries aligning with the mirror plane $m_{[010]}$ would bisect
strongly covalent bonds in the infinite
$\left[\right.$Co(C$_2$)$_2$Co$\left.\right]$ ribbons of \Co.  Thus, a
polysynthetic twin state with a stacking of twin domains along the
$a$-axis of the high-temperature phase unit cell may be proposed.

An approximate upper boundary for the twin domain sizes can be
obtained by twin integration of low-temperature single-crystal X-ray
diffraction data with \textit{EVAL14}\cite{Duisenberg03} and
subsequent data reduction with \textit{TWINABS}.\cite{Krause15}
Frame-wise scaling of collected reflection intensities as implemented
in \textit{TWINABS}\cite{Krause15} compensates for variations in
absorption and irradiated crystal volume during sample rotation. The
according scale-factors for each frame minimize intensity differences
between symmetry-equivalent reflections and can be refined separately
for each twin domain.\cite{Sheldrick07,Sheldrick15} Taking into
account the presumed polysynthetic twin domain arrangement the domain
centers in \Co\ cannot be moved to the rotation axis of the goniometer
simultaneously. As a result, large twin domains with dimensions
comparable to the X-ray beam diameter of approx. 90~$\mu$m
(FWHM)\cite{Incoatex_ImuS_AgKa} should be affected by significant and
domain-specific staggering around the rotation axis. This translates
into a different variation of symmetry-equivalent reflection
intensities for the individual twin domains and a different behavior
of the frame-wise scale-factor refined by
\textit{TWINABS}.\cite{Krause15}

\begin{figure}[h]
  \centering
  \includegraphics[width=0.8\textwidth]{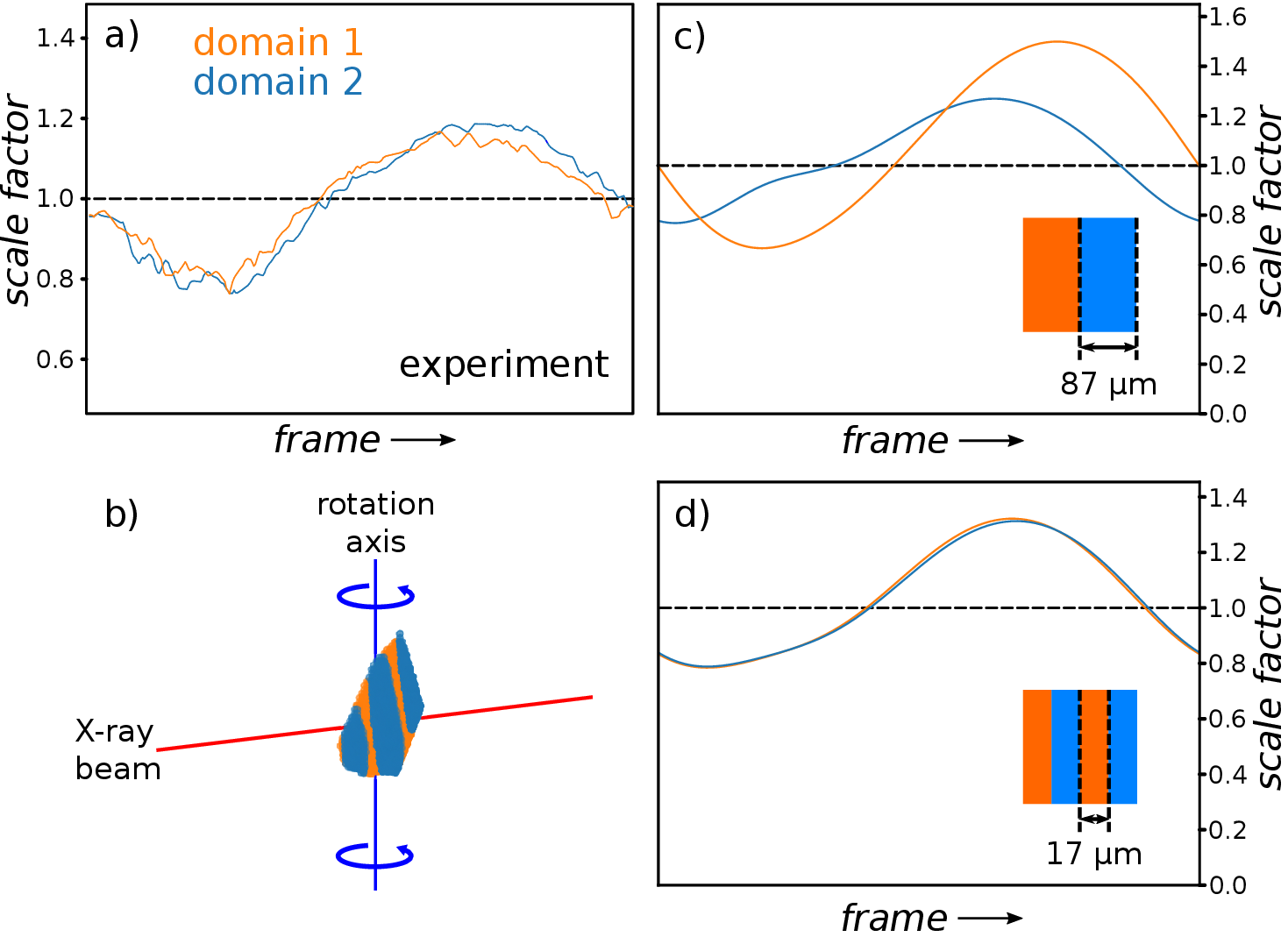}
  \caption{(a)~Frame-wise scale factor variation as obtained for a
    platelet-like crystal at 12~K and (b)~simulation setup taking into
    account the crystal shape and a Gaussian beam-profile. Simulated
    scale factor variations for a domain thickness of 87~$\mu$m and
    17~$\mu$m are given in c and d.}
  \label{fig:scale-factor}
\end{figure}

If the rotation axis is oriented perpendicular to the plane of the
domain walls, however, no differences in the domain-specific
scale-factors can be observed and a domain size assessment is not
possible. This is the case for the needle-shaped crystals whose long
axis corresponding to the crystallographic $a$-axis was oriented
approximately along the rotation axis (see crystal coordinate system
in Fig.~\ref{fig:needle}a). In contrast, the $c$-axis of a
platelet-like crystal (dimensions 140~$\mu$m~$\times$
338~$\mu$m~$\times$ 350~$\mu$m) was oriented parallel to the rotation
axis, so that differences in the scale-factor variation may be
expected in case of large twin domains. Yet, the obtained
scale-factors for the twin domains at a temperature of 12~K vary
almost perfectly in sync (see Fig.~\ref{fig:scale-factor}a). Such a
behavior points to twin domains with a thickness much smaller than the
beam diameter of 90~$\mu$m.

For a more accurate estimate of the maximum domain size the rotation
of a twinned crystal in an X-ray beam with Gaussian profile (90~$\mu$m
FWHM) was simulated. Thereby, the crystal shape was taken from the
experiment and partitioned into lamellar domains of varying thickness
(see Fig.~\ref{fig:scale-factor}b). The intensity-weighted irradiated
volumes of domains~1 and 2 were then sampled for different rotation
angles. Conversion into scale-factors was achieved by normalization of
the irradiated volumes to their average and subsequent calculation of
the reciprocal values. As can be recognized from
Figs.~\ref{fig:scale-factor}c and d, only twin domains with a
thickness below approx. 20~$\mu$m along the $a$-axis lead to a
synchronous scale-factor variation.

\newpage

\section{Generation of reciprocal-space
  maps}\label{sec:gen-rec-space-map}

For the creation of reciprocal layer reconstructions from experimental
X-ray diffraction data the program \textit{htd2predict} was
used.\cite{Langmann19} Similar to the program
\textit{XCAVATE}\cite{Estermann98} it is capable of extracting
scattering intensities in arbitrary pre-defined slices of reciprocal
space from diffraction images. Unlike precession imaging relying on a
defined specimen-alignment with respect to the rotation axis, no
pre-orientation of the sample is needed for this purpose.  The only
requirement is the use of an area detector for the acquisition of the
diffraction data. Thereby, the following procedure is used: First, the
orientation matrix describing the crystal orientation with respect to
the laboratory axis system is determined with the program
\textit{DIRAX}\cite{Duisenberg92} and refined with the program
\textit{EVAL14}.\cite{Duisenberg03} Building on this information the
positions on the diffraction images corresponding to the points in the
desired reciprocal-space layer are predicted.  The detected scattering
intensities at these positions are sampled and assigned to their
reciprocal-space coordinates.

\newpage

\section{Temperature-dependent reciprocal-space
  maps}\label{sec:rec-space-map}

\begin{figure}[h]
  \includegraphics[width=0.7\textwidth]{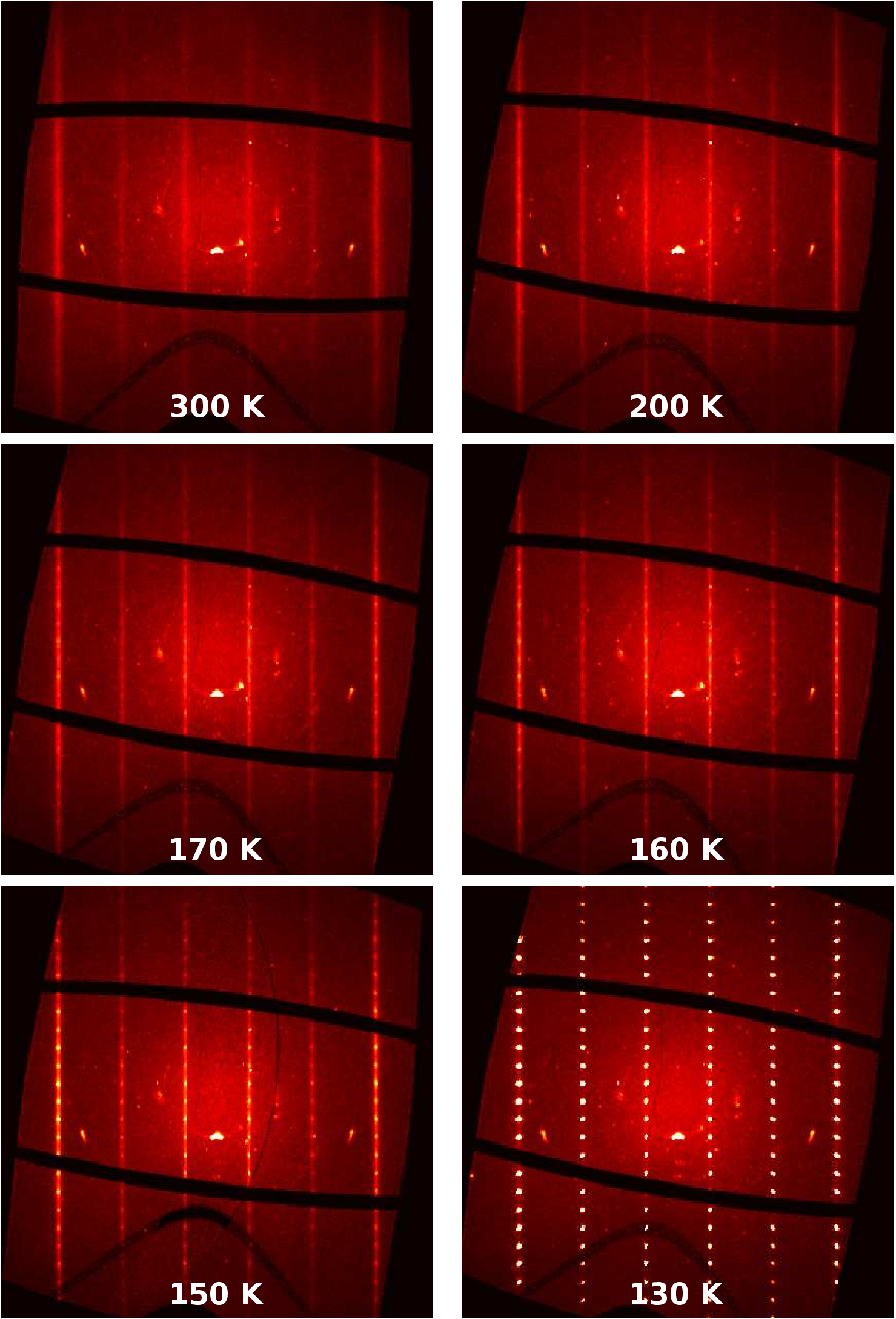}
  \caption{\label{fig:rec-space-maps-above-130K}
    Temperature-dependence of the X-ray scattering intensities in the
    $(h, 1.5, l)$ reciprocal-space plane between 300~K and 130~K.
    Note that the Miller indices refer to the orthorhombic
    high-temperature phase unit cell. The X-ray scattering intensity
    at irregular positions can be attributed to imperfections in the
    investigated large single-crystal.}
\end{figure}

\begin{figure}
  \includegraphics[width=0.7\textwidth]{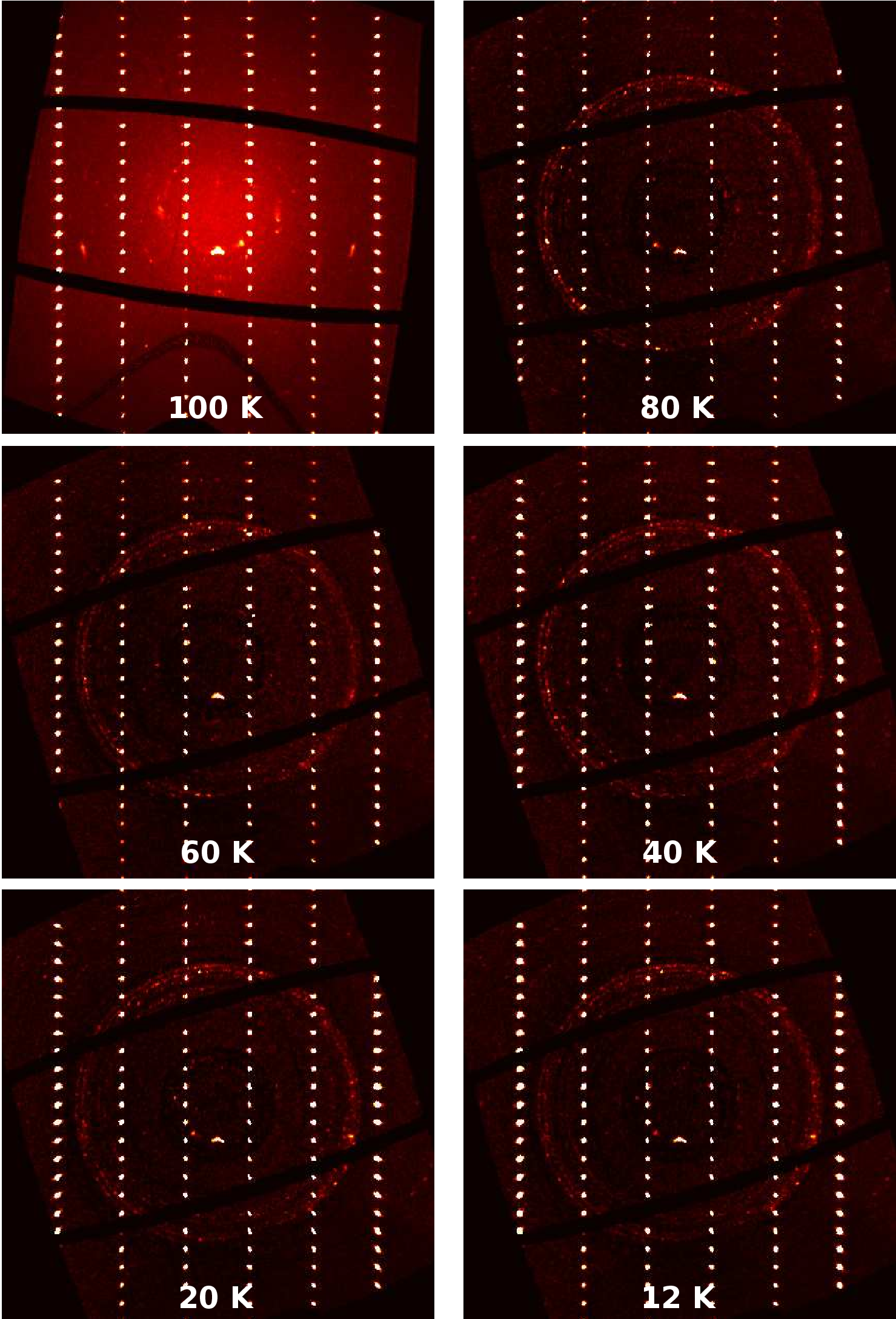}
  \caption{\label{fig:rec-space-maps-below-130K}
    Temperature-dependence of the X-ray scattering intensities in the
    $(h, 1.5, l)$ reciprocal-space plane between 100~K and 12~K. Note
    that the Miller indices refer to the orthorhombic high-temperature
    phase unit cell. The X-ray scattering intensity at irregular
    positions can be attributed to imperfections in the investigated
    large single-crystal. Weak ring-shaped intensity between 80~K and
    12~K is due to an incomplete subtraction of the scattering
    contributions from the beryllium vacuum shrouds of the
    closed-cycle He cryostat (see Sec.~\ref{sec:x-ray}).}
\end{figure}

\newpage

\section{Temperature-dependent intensity profiles}
\label{sec:profiles}

\begin{figure}[h]
  \centering
  \includegraphics[width=0.8\textwidth]{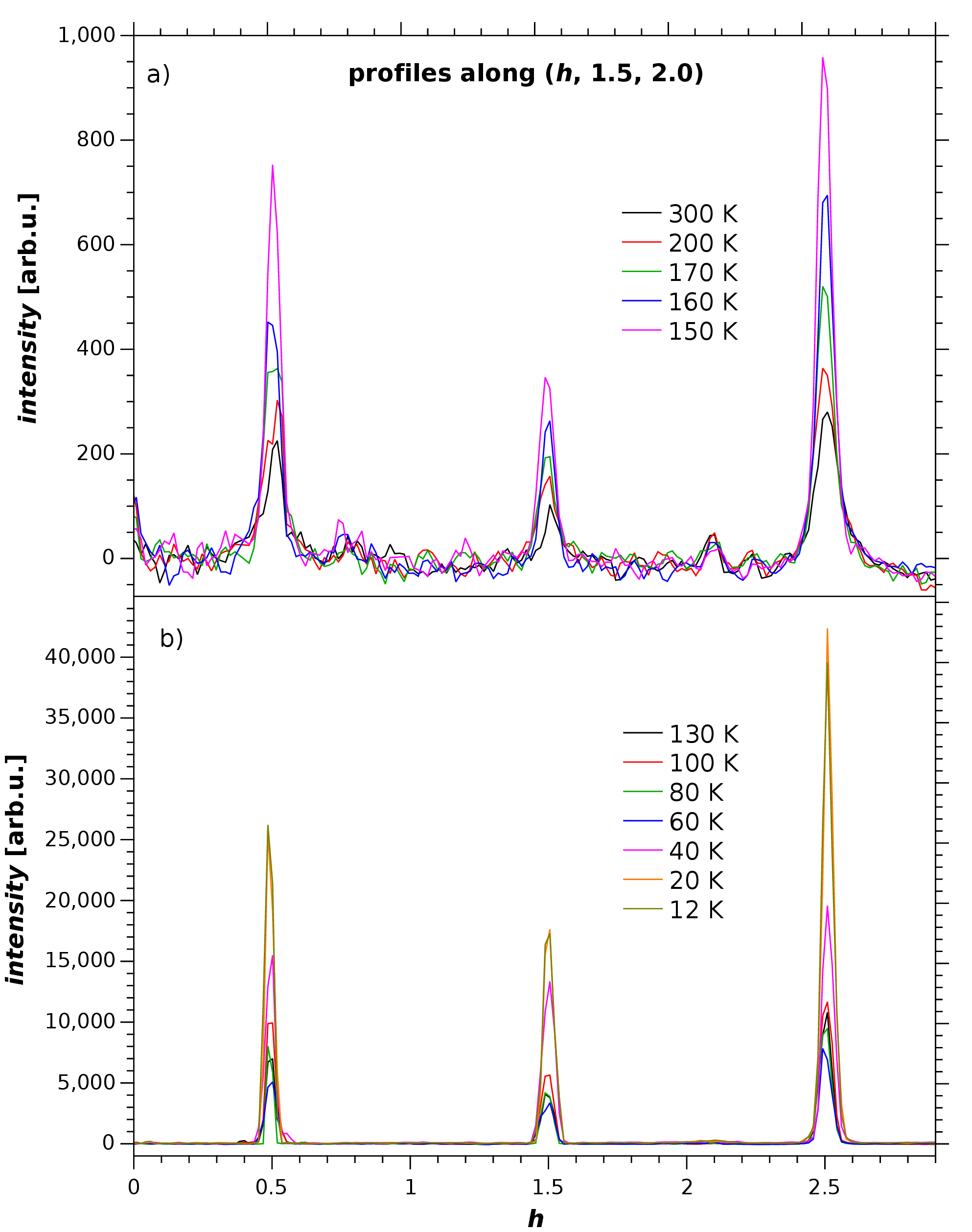}
  \caption{Horizontal cuts through the reciprocal-space maps from
    Sec.~\ref{sec:rec-space-map}. The path $(h, 1.5, 2.0)$ including
    superstructure reflection positions was sampled for temperatures
    between 300~K and 150~K (a) and between 130~K and 12~K (b). Note
    that the Miller indices refer to the orthorhombic high-temperature
    phase unit cell and that a Gaussian background was subtracted.}
  \label{fig:profile_at_1}
\end{figure}

\begin{figure}[p]
  \centering
  \includegraphics[width=0.8\textwidth]{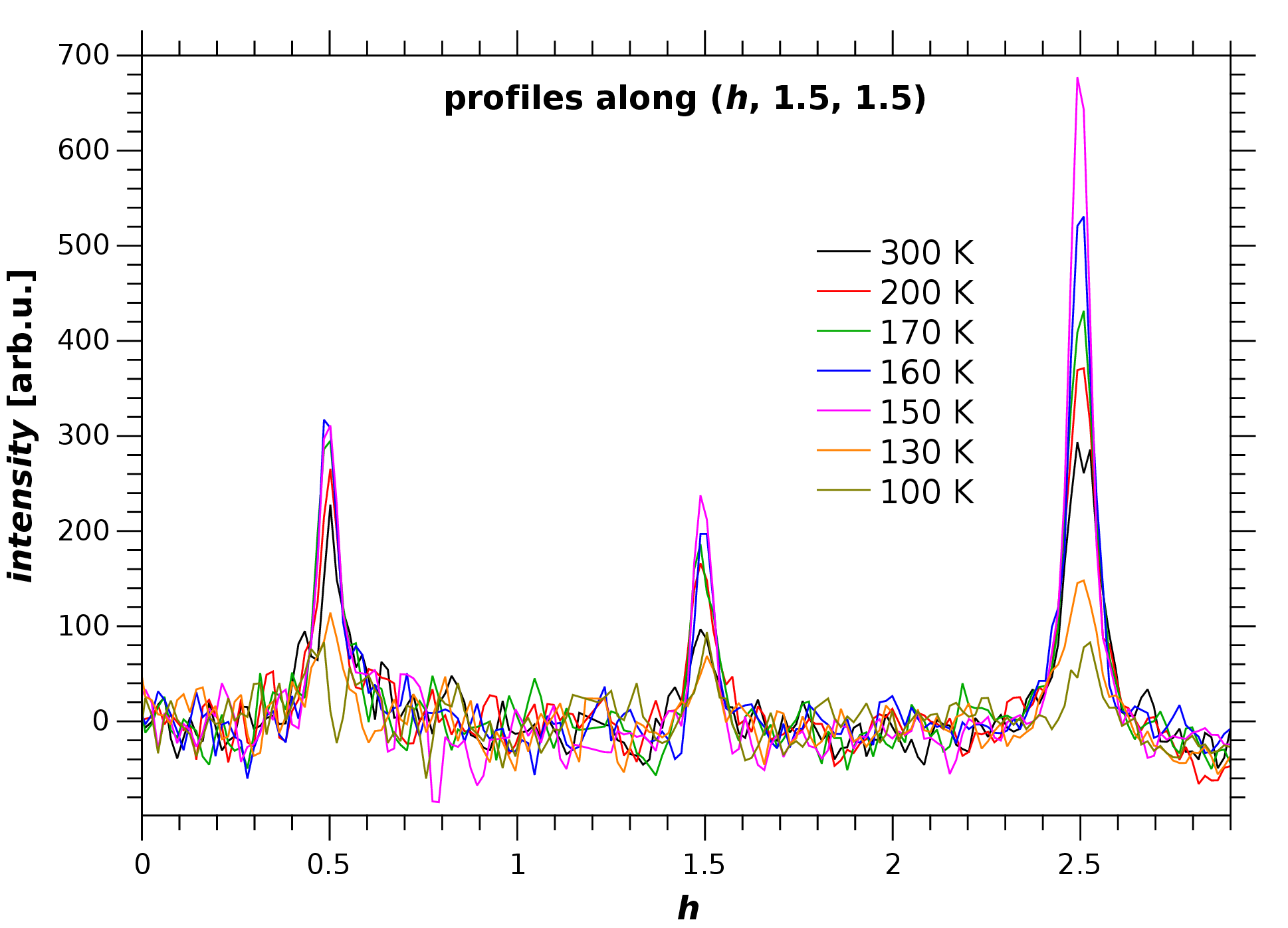}
  \caption{Horizontal cuts through the reciprocal-space maps from
    Sec.~\ref{sec:rec-space-map}. The path $(h, 1.5, 1.5)$ excluding
    superstructure reflection positions was sampled for temperatures
    between 300~K and 100~K. Note that the Miller indices refer to the
    orthorhombic high-temperature phase unit cell and that a Gaussian
    background was subtracted.}
  \label{fig:profile_at_2}
\end{figure}

\FloatBarrier

\newpage

\section{Extraction of scattering intensities}
\label{sec:extr-scatt-intens}

Due to different characteristics of diffuse rods and superstructure
reflections adapted approaches had to be used in the extraction of
their intensities.

The superstructure reflections are pinpoint and spread only over few
X-ray diffraction images collected in $\phi$-scanning mode.  This
situation allows the usage of single unprocessed diffraction images to
track the temperature-dependent intensity of a representative
$(-1.5, 0.5, 0)$ reflection.\footnote{referring to the unit cell of
  the high-temperature phase} The intensity sampling was done using
the program Imagej\cite{Schneider12} by summing up the detected
intensity of the pixels inside a quadratic box around the predicted
position of the $(-1.5, 0.5, 0)$ reflection. Therefrom, a background
intensity obtained by multiplication of the average pixel intensity at
the edges of the integration box with the number of contained pixels
was subtracted. The usage of different $\phi$-increments of
0.5$^\circ$ and 2.0$^\circ$ per frame in the temperature ranges
10~K~$< T <$ 100~K and 100~K~$< T <$ 300~K was outbalanced by adding
up four frames with 0.5$^\circ$ increment. Both temperature-dependent
intensity data sets were scaled to the reflection intensity at 100~K.

The diffuse rods, by contrast, are weak and span a large number of
X-ray diffraction images. Thus, the intensity extraction scheme was
based on reciprocal-space reconstructions that were generated with the
program \textit{htd2predict}\cite{Langmann19} (see
Sec.~\ref{sec:gen-rec-space-map}). Thereby, the intensity at positions
$(2.5, 1.5, 0.5)$, $(2.5, 1.5, 1.5)$ and
$(2.5, 1.5, 2.5)$,\cite{Note1} \textit{i.e.}  midway between the
superstructure reflection positions in a row along $c^\ast$, was
sampled by a box integration method in analogy to the superstructure
reflections. Due to their continuous nature along $c^\ast$, however,
only the pixel intensity at the two edges of the integration box
parallel to the diffuse rods was used in the determination of the
subtracted background intensity. In a final step the intensities at
all three positions were averaged.

\newpage

\section{Simulations of thermal diffuse scattering (TDS)}
\label{sec:TDS-sim}

In this work the simulation of thermal diffuse scattering (TDS)
contributions with the program \textit{ab2tds}\cite{Wehinger13} relies
on \textit{ab-initio} dynamical matrices obtained by the
finite-displacement method. Necessary Fourier interpolation of the
dynamical matrix\cite{Wehinger14a} during the simulation process may
introduce artifacts such as the observed weak modulation of the
diffuse rod intensity for the high-temperature (HT) phase of \Co\
along the $c^\ast$-axis.

To ensure that the dynamical matrix is basically left intact by the
Fourier interpolation step inelastic X-ray scattering (IXS) intensity
maps for a temperature of 300~K were simulated in
\textit{ab2tds}.\cite{Wehinger13} By plotting the calculated variation
of the inelastically scattered X-ray intensity $I(E, \mathbf{Q})$ with
energy transfer $E$ and momentum transfer $\mathbf{Q}$ salient
features of the phonon dispersion should be reproduced.  Differences
may only arise due to the dependence of $I(E, {\bf Q})$ on the
scattering factors $S_\alpha(\lambda, \mathbf{Q})$ for atoms $\alpha$
and the scalar product
$\mathbf{Q} \cdot \mathbf{e}^\alpha_{v\mathbf{Q}}$ of momentum
$\mathbf{Q}$ and polarisation vector for phonon branch $v$.
Additionally, folding of $I(E, {\bf Q})$ with a resolution function
leads to a blurring of details in the phonon dispersion.

\begin{figure}[h]
  \centering
  \includegraphics[width=0.9\textwidth]{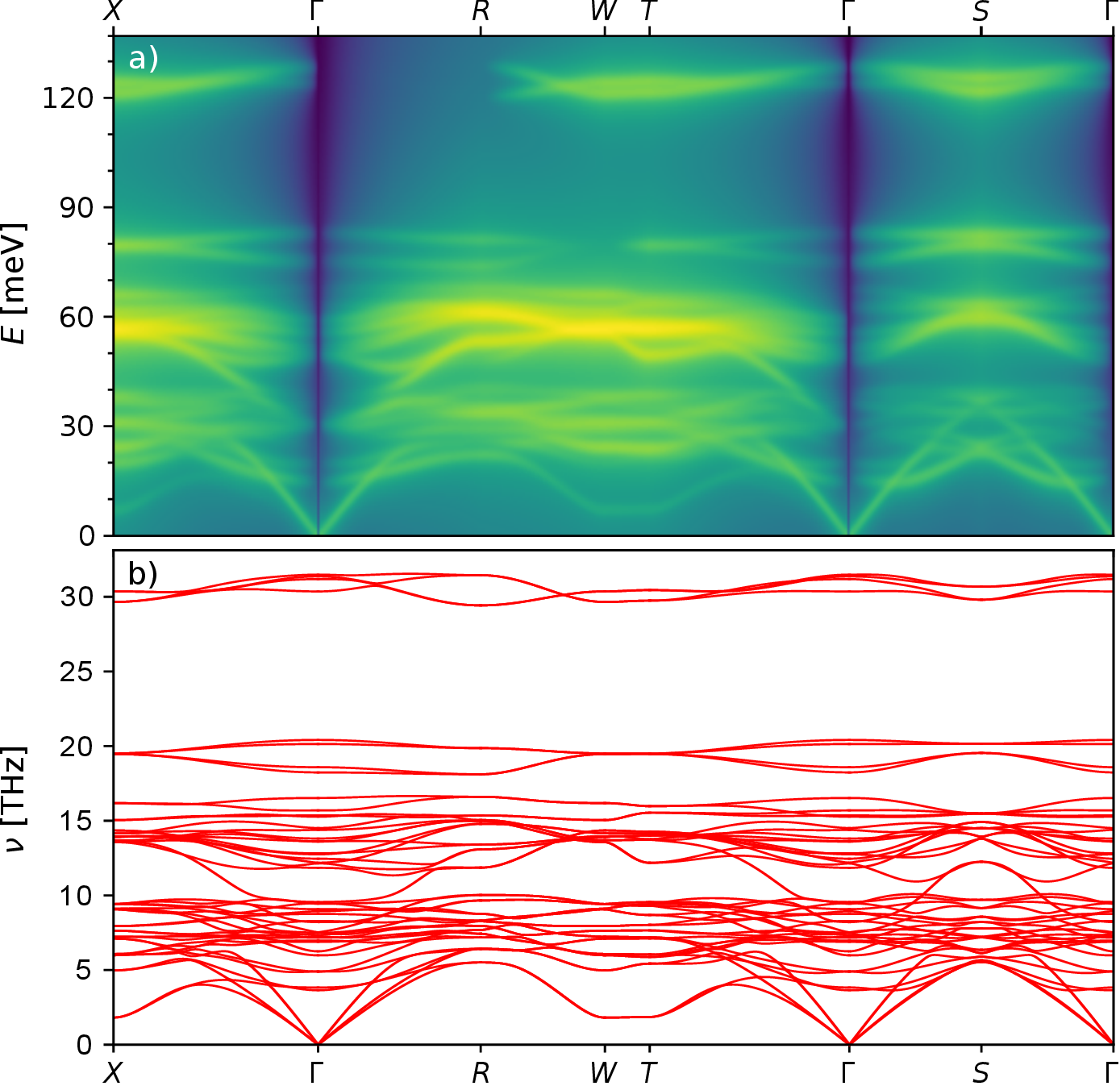}
  \caption{(a)~Simulated Inelastic X-ray scattering (IXS) map at 300~K
    and (b)~unprocessed phonon dispersion relation for the
    high-temperature phase of \Co\ along the same high-symmetry paths
    in the Brillouin zone.}
  \label{fig:tds-dispers-ht}
\end{figure}

\begin{figure}[h]
  \centering
  \includegraphics[width=0.9\textwidth]{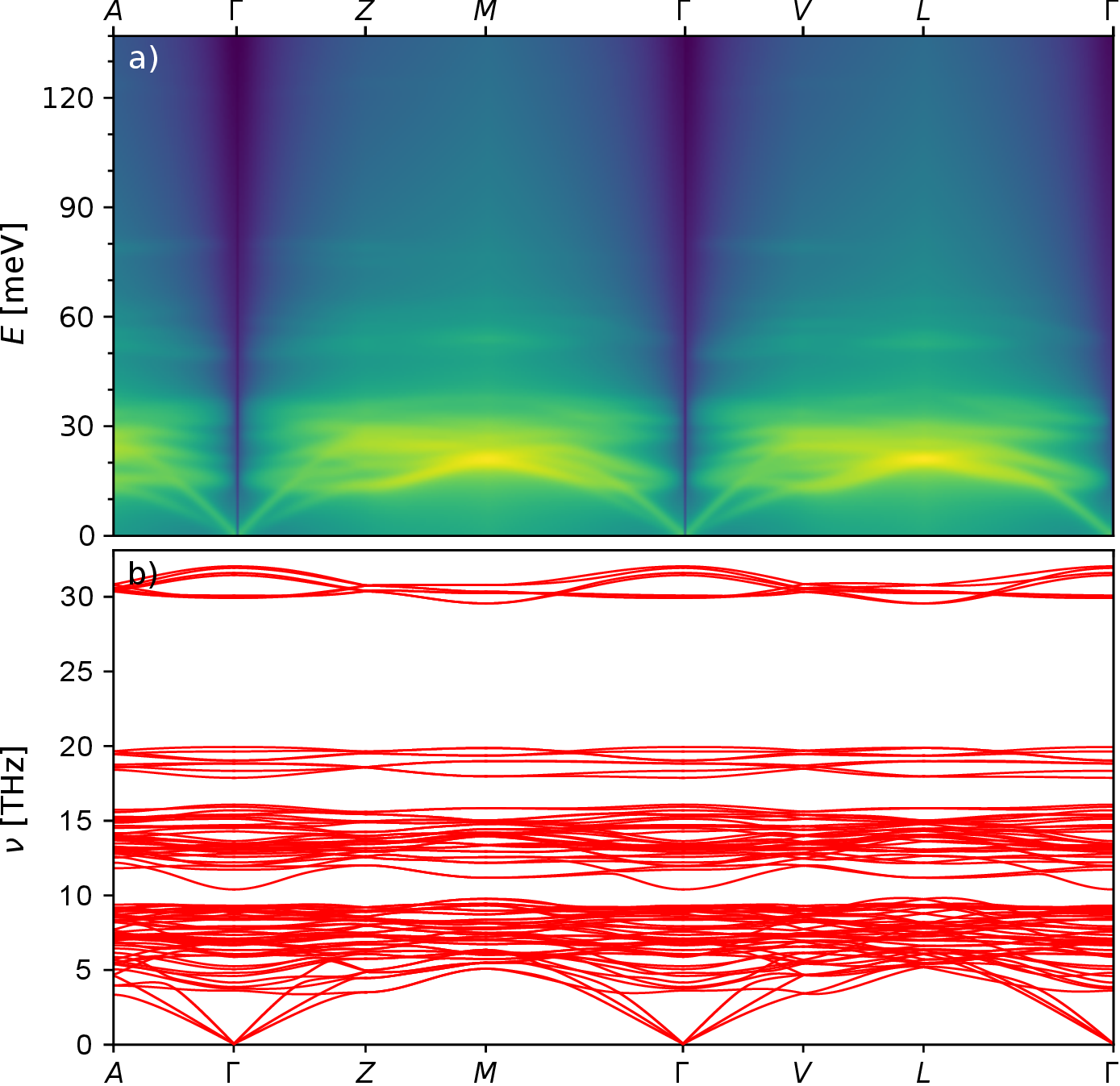}
  \caption{(a)~Simulated Inelastic X-ray scattering (IXS) map at 300~K
    and (b)~unprocessed phonon dispersion relation for the
    low-temperature phase of \Co\ along the same high-symmetry paths
    in the Brillouin zone.}
  \label{fig:tds-dispers-lt}
\end{figure}

Comparison of the simulated IXS intensity map for the HT-phase of \Co\
in Fig.~\ref{fig:tds-dispers-ht}a with the unprocessed phonon
dispersion in Fig.~\ref{fig:tds-dispers-ht}b shows excellent overall
agreement. Most noteworthy, the soft phonon branch between the
high-symmetry points W and T in the phonon dispersion reappears
unaltered in the IXS intensity map. Vanishing of the IXS intensity for
the phonon branches between $\Gamma$ and R at approx. 120~meV may be
caused by a small $S_\alpha(\lambda, \mathbf{Q})$ or orthogonality of
$\mathbf{Q}$ and $\mathbf{e}^\alpha_{v\mathbf{Q}}$ in this
region. There is also good correspondence between the simulated IXS
intensity map for the LT-phase of \Co\ in
Fig.~\ref{fig:tds-dispers-lt}a and the unprocessed phonon dispersion
in Fig.~\ref{fig:tds-dispers-lt}b. Although high-energy regions of the
IXS map suffer from weak intensity, the absence of a soft phonon
branch in the LT-phase is clearly reflected.

\FloatBarrier

\bibliography{SupportingInfo.bib}